\documentclass[conference]{IEEEtran}
\makeatletter
\def\ps@headings{%
\def\@oddhead{\mbox{}\scriptsize\rightmark \hfil \thepage}%
\def\@evenhead{\scriptsize\thepage \hfil \leftmark\mbox{}}%
\def\@oddfoot{}%
\def\@evenfoot{}}
\makeatother 
 \IEEEoverridecommandlockouts

\usepackage{epsfig, latexsym, amssymb, amsmath,url}
\usepackage{graphics}
\usepackage{algorithm}
\usepackage{algorithmic}
\usepackage{subfigure}

\newcommand{\tabincell}[2]{\begin{tabular}{@{}#1@{}}#2\end{tabular}}

\newtheorem{prop}{Proposition}[section]

\newtheorem{cor}{Corollary}

\newtheorem{lm}{Lemma}

\newtheorem{thm}{Theorem}

\newcommand{\bthm}{\begin{thm}}
\newcommand{\ethm}{\end{thm}}

\newcommand{\bcor}{\begin{cor}}
\newcommand{\ecor}{\end{cor}}
\newcommand{\bprop}{\begin{prop}}
\newcommand{\eprop}{\end{prop}}
\newcommand{\blm}{\begin{lm}}
\newcommand{\elm}{\end{lm}}
\newcommand{\beq}{\begin{equation}}
\newcommand{\eeq}{\end{equation}}
\newcommand{\ber}{\begin{eqnarray}}
\newcommand{\eer}{\end{eqnarray}}

\newenvironment{proof1}{\begin{trivlist}\item[]{\bf Proof:\hspace{2mm}}}{\hfill$\blackbox$\end{trivlist}}
\newcommand{\bproof}{\begin{proof1}}
\newcommand{\eproof}{\end{proof1}}

\newcommand{\blackbox}{\vrule height7pt width5pt depth1pt}

\newcommand{\bit}{\begin{itemize}}
\newcommand{\eit}{\end{itemize}}
\newcommand{\ben}{\begin{enumerate}}
\newcommand{\een}{\end{enumerate}}
\newcommand{\bdesc}{\begin{description}}
\newcommand{\edesc}{\end{description}}
\newcommand{\beqarrn}{\begin{eqnarray*}}
\newcommand{\eeqarrn}{\end{eqnarray*}}
\newenvironment{proofof}[1]{\begin{trivlist}\item[]{\bf Proof of #1:\hspace{2mm}
}}{\hfill\blackbox\end{trivlist}}
\newcommand{\bproofof}{\begin{proofof}}
\newcommand{\eproofof}{\end{proofof}}
\newenvironment{rem}{\begin{trivlist}\item[]{\bf
Remark:}\hspace{4mm}}{\end{trivlist}}
\newcommand{\brem}{\begin{rem}}
\newcommand{\erem}{\end{rem}}
\newenvironment{rems}{\begin{trivlist}\item[]{\bf
Remarks}\begin{itemize}}{\end{itemize}\end{trivlist}}
\newcommand{\brems}{\begin{rems}}
\newcommand{\erems}{\end{rems}}
\newtheorem{fact}{Fact}
\newcommand{\bfact}{\begin{fact}}
\newcommand{\efact}{\end{fact}}
\newtheorem{examp}{Example}
\newcommand{\bexamp}{\begin{examp}\rm}
\newcommand{\eexamp}{\end{examp}}
\newtheorem{defn}{Definition}
\newcommand{\bdefn}{\begin{defn}\rm}
\newcommand{\edefn}{\end{defn}}

\newtheorem{prob}{Problem}
\newcommand{\bprob}{\begin{prob}}
\newcommand{\eprob}{\end{prob}}

\newcommand{\bvtm}{\begin{verbatim}}
\newcommand{\bfig}{\begin{figure}}
\newcommand{\efig}{\end{figure}}
\newcommand{\bcen}{\begin{center}}
\newcommand{\ecen}{\end{center}}

\long\def\comment#1{}

\def \n2{{N_0 \over 2}}

\def \h5{\hspace{0.5in}}

\begin{document}

\title{Optimal Allocation of Interconnecting Links in Cyber-Physical Systems: Interdependence, Cascading Failures and Robustness}


\author{
Osman Ya\u{g}an, Dajun Qian,  Junshan Zhang, and Douglas Cochran\\
{\tt \{oyagan, dqian, junshan.zhang, cochran\}}{\tt @asu.edu}
 \\
School of Electrical, Computer and Energy Engineering \\
Arizona State University, Tempe, AZ 85287-5706 USA\\
}

%


\maketitle
 \pagestyle{empty}
  \thispagestyle{empty}

\begin{abstract}

We consider a cyber-physical system consisting of two interacting
networks, i.e., a cyber-network overlaying a physical-network. It
is envisioned that these systems are more vulnerable to attacks
since node failures in one network may result in (due to the
interdependence) failures in the other network, causing a cascade
of failures that would potentially lead to the collapse of the
entire infrastructure. The robustness of interdependent systems
against this sort of catastrophic failure hinges heavily on the
allocation of the (interconnecting) links that connect nodes in
one network to nodes in the other network. In this paper, we
characterize the {\em optimum} inter-link allocation strategy
against random attacks in the case where the topology of each
individual network is  unknown. In particular, we analyze the \lq
\lq regular" allocation strategy that allots exactly the same
number of bi-directional inter-network links to all nodes in the
system. We show, both analytically and experimentally, that this
strategy yields  better performance (from a network resilience
perspective) compared to all possible strategies, including
strategies using random allocation, unidirectional inter-links,
etc.
\end{abstract}


\section{Introduction}

Today's worldwide network infrastructure consists a web of
interacting cyber-networks (e.g., the Internet) and physical
systems (e.g., the power grid). There is a consensus that
integrated cyber-physical systems will emerge as the underpinning
technology for major industries in the 21st century \cite{CPS}.
The smart grid is one archetypal example of such systems where the
power grid network and the communication network for its
operational control are coupled together and depend on each other;
i.e., they are {\em interdependent}. While interdependency allows
building systems that are larger, smarter and more complex, it has
been observed \cite{Vespignani} that interdependent systems tend
to be more fragile against failures, natural hazards and attacks.
For example, in the event of an attack to an interdependent
system, the failures in one of the networks can cause failures of
the dependent nodes in the other network and vice versa. This
process may continue in a recursive manner and hence lead to a
cascade of failures causing a catastrophic impact on the overall
cyber-physical system. In fact, the cascading effect of even a
partial Internet blackout could disrupt major national
infrastructure networks involving Internet services, power grids
and financial markets \cite{Buldyrev}. Real-world examples include
the $2003$ blackout in the northeastern United States and
southeastern Canada \cite{Vespignani} and the electrical blackout
that affected much of Italy on 28 September 2003 \cite{Buldyrev}.

\subsection{Background and Related Work}
Despite recent studies of cascading failures in complex networks,
the dynamics of such failures and the impact  across multiple
networks are not well understood. There is thus a need to develop
a new network science for modeling and quantifying cascading
failures, and to develop network management algorithms that
improve network robustness and ensure overall network reliability
against cascading failures. Most existing studies on failures in
complex networks consider only the single network case. A notable
exception is the very recent work of Buldyrev et al.
\cite{Buldyrev} in which a \lq \lq one-to-one correspondence"
model for studying the ramifications of interdependence between
two networks is set forth. This model considers two networks of
the same size, say network $A$ and network $B$, where each node in
network $A$ depends on one and only one node in network $B$ and
vice versa. In other words, each node in network $A$ has one
bi-directional {\em inter-edge} connecting it to a {\em unique}
node in network $B$. Furthermore, it is assumed that a node in
either network can function {\em only if} it has support from the
other network; i.e., it is connected (via an inter-edge) to at
least one functioning node from the other network.

The robustness of the one-to-one correspondence model was studied
in \cite{Buldyrev} using a similar approach to that of the works
considering single networks
\cite{callaway2000network,cohen2000resilience}. Specifically, it
is assumed that a random attack is launched upon network $A$,
causing the failure of a fraction $1-p$ of the nodes; this was
modeled by a random removal of a fraction $1-p$ of the nodes from
network $A$. Due to the interdependency, these initial failures
lead to node failures from network $B$, which in turn may cause
further failures from network $A$ thereby triggering an avalanche
of cascading failures. To evaluate the robustness of the model,
the size of the functioning parts of both networks are computed at
each stage of the cascading failure until a {\em steady state} is
reached; i.e., until the cascade of failure ends. One of the
important findings of \cite{Buldyrev} was to show the existence of
a critical threshold on $p$, denoted by $p_c$, above which a
considerable fraction of nodes in both networks remain functional
at the steady state; on the other hand, if $p<p_c$, both networks
go into a complete fragmentation and the entire system collapses.
Also, it is observed in \cite{Buldyrev} that interdependent
network systems have a much larger $p_c$ compared to that of the
individual constituent networks; this is compatible with the
observation that interdependent networks are more vulnerable to
failures and attacks.

The original work of Buldyrev et al. \cite{Buldyrev} has received
much attention and spurred the study of interdependent networks in
many different directions; e.g., see
\cite{BuldyrevShereCwilich,ChoGohKim,HuangGaoBuldyrevHavlinStanley,
ParshaniBuldyrevHavlin,SchneiderArajuoHavlinHerrman,ShaoBuldyrevHavlinStanley}.
One major vein of work, including \cite{BuldyrevShereCwilich,
ParshaniBuldyrevHavlin,ShaoBuldyrevHavlinStanley}, aims to extend
the findings of \cite{Buldyrev} to more realistic scenarios than
the one-to-one correspondence model. More specifically, in
\cite{BuldyrevShereCwilich} the authors consider a one-to-one
correspondence model with the difference that mutually dependent
nodes are now assumed to have the same number of neighbors in
their own networks; i.e., their intra-degrees are assumed to be
the same. In \cite{ParshaniBuldyrevHavlin} the authors consider
the case where only a fraction of the nodes in network $A$ depend
on the nodes in network $B$, and vice versa. In other words, some
nodes in one network are assumed to be {\em autonomous}, meaning
that they do not depend on nodes of the other network to function
properly. Nevertheless, in \cite{ParshaniBuldyrevHavlin} it was
still assumed that a node can have {\em at most} one supporting
node from the other network. More recently, Shao et al.
\cite{ShaoBuldyrevHavlinStanley} pointed out the fact that, in a
realistic scenario, a node in network $A$ may depend on more than
one node in network $B$, and vice versa. In this case, a node will
function as long as at least one of its supporting nodes is still
functional. To address this case, Shao et al.
\cite{ShaoBuldyrevHavlinStanley} proposed a model where the
inter-edges are unidirectional and each node supports (and is
supported by) a {\em random} number of nodes from the other
network. In a different line of work, Schneider et al.
\cite{SchneiderArajuoHavlinHerrman} adopted a design point of view
and explored ways to improve the robustness of the one-to-one
correspondence model by letting some nodes be autonomous. More
precisely, they assume that the topologies of networks $A$ and $B$
are known and propose a method, based on degree and centrality,
for choosing the autonomous nodes properly in order to maximize
the system robustness.

\subsection{Summary of Main Results}
In this study, we stand in the intersection of the two
aforementioned lines of work. First, we consider a model where
inter-edges are allocated {\em regularly} in the sense that all
nodes have exactly the {\em same} number of {\em bi-directional}
inter-edges, assuming that no topological information is
available. This ensures a uniform support-dependency relationship
where each node supports (and is supported by) the same number of
nodes from the other network. We analyze this new model in terms
of its robustness against random attacks via characterizing the
steady state size of the functioning parts of each network as well
as the critical fraction $p_c$. In this regard, our work
generalizes the studies on the one-to-one correspondence model and
the model studied by Shao et al. \cite{ShaoBuldyrevHavlinStanley}.
From a design perspective, we show analytically that the proposed
method of {\em regular} inter-edge allocation improves the
robustness of the system over the random allocation strategy
studied in \cite{ShaoBuldyrevHavlinStanley}. Indeed, for a given
{\em expected} value of inter-degree (the number of nodes it
supports {\em plus} the number of nodes it depends upon) per node,
we show that: i) it is better (in terms of robustness) to use
bi-directional inter-links than unidirectional links, and ii) it
is better (in terms of robustness) to deterministically allot each
node exactly the same number of bi-directional inter-edges rather
than allotting each node a random number of inter-edges.

These results imply that if the topologies of network $A$ and
network $B$ are unknown, then the optimum inter-link allocation
strategy is to allot exactly the same number of bi-directional
inter-edges to all nodes. Even if the statistical information
regarding the networks is available; e.g., say it is known that
network $A$ is an Erd\H{o}s-R\'enyi \cite{Bollobas} network and
network $B$ is a scale-free network \cite{BarabasiAlbert}, regular
inter-edge allocation is still the best strategy in the absence of
the detailed topological information; e.g., in the case where it
is not possible to estimate the nodes that are likely to be more
important in preserving the connectivity of the networks, say
nodes with high betweenness \cite{CohenHavlin}. Intuitively, this
makes sense because without knowing which nodes play a key role in
preserving the connectivity of the networks, it is best to treat
all nodes \lq\lq identically" and give them equal priority in
inter-edge allocation.

The theoretical results in this paper are also supported by
extensive computer simulations. Numerical results are given for
the case where both networks are Erd\H{o}s-R\'enyi (ER) and the
optimality of the regular allocation strategy is verified. To get
a more concrete sense, assume that $A$ and $B$ are ER networks
with $N$ nodes and average degree $4$. When inter-edges are
allocated regularly so that each node has exactly $2$
bi-directional inter-edges, the critical threshold $p_c$ is equal
to $0.43$. However, for the same networks $A$ and $B$, if the
number of inter-edges follows a Poisson distribution with mean
$2$, the critical $p_c$ turns out to be equal to $0.82$. This is a
significant difference in terms of robustness, since in the former
case the system is resilient to the random failure of up to $57
\%$ of the nodes while in the latter case, the system is resilient
to the random failure of up to only $18 \%$ of the nodes.

To the best of our knowledge, this paper is the first work that
characterizes the robustness of interdependent networks under
regular allocation of bi-directional inter-edges. Also, it is the
first work that determines analytically and experimentally the
optimum inter-edge allocation strategy in the absence of
topological information. We believe that our findings along this
line shed light on the design of interdependent systems.

\subsection{Structure of the Paper}
The paper is organized as follows. In Section \ref{sec:Model}, we
introduce the system model and present an overview of cascading
failures. The behavior of this model under random attacks is
analyzed in Section \ref{sec:analysis}, where  the functional size
of the networks is characterized. Section \ref{sec:proof of
regular} is devoted to proving the optimality of regular
inter-link allocation, while in Section \ref{sec:Numerical} we
give numerical examples and simulation results. Possible future
research is explained and the paper is concluded in Section
\ref{sec:conclusion}.

\section{System Model}\label{sec:Model}
We consider a cyber-physical system consisting of two interacting
networks, say network $A$ and network $B$. For simplicity, both
networks are assumed to have $N$ nodes and the vertex sets in
their respective graphical representations are denoted by
$\{v_1,\ldots,v_N\}$ and $\{v_1',\ldots,v_N'\}$. We refer to the
edges connecting nodes within the same network as {\em
intra-edges} and those connecting nodes from two different
networks as {\em inter-edges}. Simply put, we assume that a node
can function {\em only if} it is connected (via an inter-edge) to
at least one functioning node in the other network
\cite{Buldyrev}; and we will elaborate further on this. Clearly,
the interdependency between two networks is intimately related to
the inter-edges connecting them. In this study,
 inter-edges are assumed to be bi-directional
so that it is convenient to use an $N \times N$ interdependency
matrix $\mathbf{C}$ to represent the bi-directional inter-edges
between networks $A$ and $B$. Specifically, for each
$n,m=1,\ldots,N$, let
\begin{equation}
\left(\mathbf{C}\right)_{nm} = \left \{
\begin{array}{ll}
1 & \mbox{if~$v_n$~and~$v_m'$~depend~on~each~other} \\
0 & \mbox{otherwise}
\end{array}
\right . \label{eq:C}
\end{equation}

We also assume that inter-edges are allocated {\em regularly} so
that each node has exactly $k$ inter-edges, where $k$ is  an
integer satisfying $k \leq N$. Without loss of generality, this
strategy can be implemented in the following manner: For each
$n=1,2,\ldots,N$, let the interdependency matrix be given by
\begin{equation}
\left(\mathbf{C}\right)_{nm} = \left \{
\begin{array}{ll}
1 & \mbox{if~~~$
  m=n,n \oplus 1, \ldots, n \oplus (k-1) $} \\
0 & \mbox{otherwise},
\end{array}
\right. \label{eq:alpha_N_establish}
\end{equation}
where we define
\[
n \oplus l = \left \{
\begin{array}{ll}
n+\ell & \mbox{if~~~$n + l \leq N$} \\
n+\ell - N & \mbox{if~~~$n + l > N$},
\end{array}
\right.
\]
for each $\ell=0,1,\ldots, N-1$; see also Figure \ref{fig:set-up}.

\begin{figure}[!t]
\begin{center}
\includegraphics[totalheight=0.3\textheight,
width=.6\textwidth]{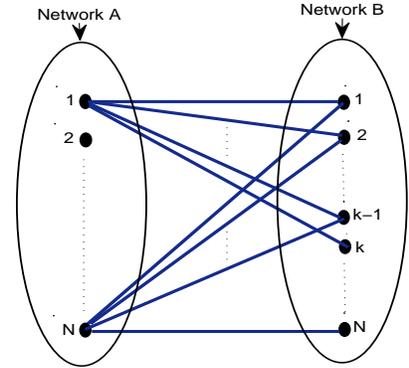}\caption{\sl A sketch of the
proposed system model, namely the {\em regular} allocation
strategy of bi-directional inter-edges: Each node in $A$ is
connected to exactly $k$ nodes in $B$, and vice versa.}
\label{fig:set-up}
\end{center}
\end{figure}

\begin{figure*}[!ht]
\includegraphics[totalheight=0.18\textheight,
width=.82\textwidth]{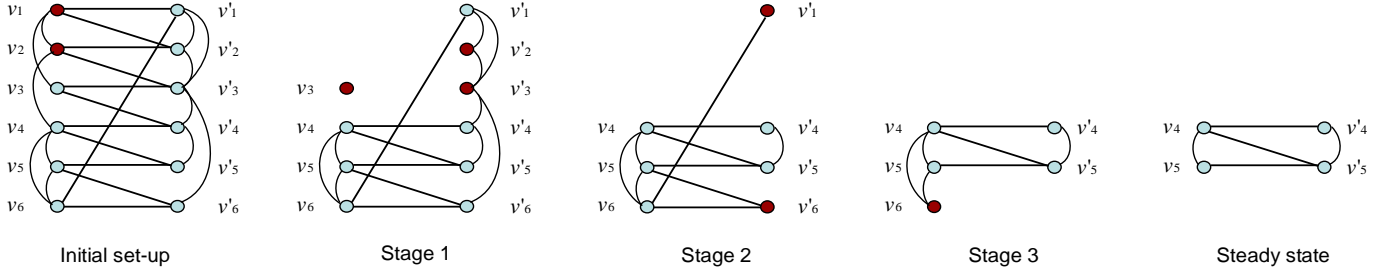}\caption{\sl An illustration of
cascading failures in two interdependent networks. Network $A$
with nodes $\{v_1,v_2,\ldots, v_6\}$ and network $B$ with nodes
$\{v'_1,v'_2,\ldots, v'_6\}$ are interdependent with each node
having exactly two bi-directional inter-edges. Initially, a random
attack causes the failure of nodes $v_1$ and $v_2$. In stage $1$,
$v_1$ and $v_2$ are removed from the system along with all the
links (inter and intra) that are incident upon them. As a result,
node $v_3$ becomes disconnected from the functioning giant
component of network $A$, and thus fails. These failures then
cause the nodes $v_2'$ and $v_3'$ to fail as they lose all their
supports; i.e., all the inter-edges that are incident upon them
are removed. In stage $2$, we see the effect of removing $v_2'$
and $v_3'$ from network $B$: nodes $v_1'$ and $v_6'$ fail as they
become disconnected from the functioning giant component. The
failure of nodes $v_1'$ and $v_6'$ then leads to the failure of
node $v_6$ in stage $3$, since $v_6$ was being supported solely by
$v_1'$ and $v_6'$. By removal of the node $v_6$, the failures stop
and the system reaches steady state.} \label{fig:demo}
\end{figure*}

We are interested in evaluating the network robustness in the case
of random node failures (or equivalently random attacks).
Specifically, in the dynamics of cascading failures, we assume
that a node is {\em functioning} at Stage $i$ if the following
conditions are satisfied
\cite{Buldyrev,ShaoBuldyrevHavlinStanley}: i) The node has at
least one inter-edge with a node that was functioning at Stage
$i-1$; ii) The node belongs to the giant (i.e., the largest)
component of the sub-network formed by the nodes (of its own
network) that satisfy condition i). For both networks, a giant
component consisting of functioning nodes will be referred to as a
{\em functioning giant component}.

We assume that the cascade of failures is triggered by the failure
of a fraction $1-p$ of the nodes in network $A$. We further assume
that these $(1-p)N$ nodes are chosen (say by the attacker)
uniformly at random amongst all nodes in network $A$. By the
definitions given above, it can be seen that after the initial
attack, only nodes in the functioning giant component of $A$ can
operate properly. As a result of that, in the next stage, some of
the nodes in network $B$ may end up losing all of their
inter-connections and turn dysfunctional. In that case, the nodes
that can function properly in network $B$ will only be those in
the functioning giant component of $B$. But, this fragmentation of
network $B$ may now trigger further failures in network $A$ due to
nodes that lose all their $B$-connections. Continuing in this
manner, the cascade of failures propagates alternately between $A$
and $B$, eventually (i.e., in steady state) leading to either:
$1)$ {\em residual functioning giant components in both networks},
or $2)$ {\em complete failure of the entire system}. For an
illustrative example, see Figure \ref{fig:demo} where a cascading
failure is demonstrated for a pair of interdependent networks with
$N=6$ nodes, $k=2$, and $p=2/3$.

\section{Analysis of Cascading Failures under Regular Allocation of Inter-edges}
\label{sec:analysis}

In this section, we analyze the dynamics of cascading failures in
two interacting networks. A principal objective of this study is
to quantify the effectiveness of the regular allocation strategy
for network robustness, by means of: i) characterizing the size of
the remaining giant components in networks $A$ and $B$ after the
cascade has reached a steady state, and ii) finding the
corresponding critical threshold $p_c$. To these ends, we will use
the technique of generating functions
\cite{newman2002spread,newman2001random} to analyze the sizes of
functioning giant components in the two networks at each stage.
For convenience, the notation used in the calculations is
summarized in Table ~\ref{tb:notation}. {\small
\begin{table}{}
 \caption{ Key notation  in the analysis of cascading
failures}
\begin{center}
\begin{tabular}{|l|l| }
\hline  $A_i$, $B_i$  & the  functioning giant components in A and B at stage $i$\\
\hline   $p_{Ai}$, $p_{Bi}$  & \tabincell{l} { the fractions
corresponding to
functioning giant components \\
at stage $i$,  $|A_i|=p_{Ai}  N$, $|B_i|=p_{Bi}  N$                     }\\
\hline $\bar A_i $, $\bar B_i$  & \tabincell{l}{ the remaining
nodes in $A$ and  $B$ retaining at  least one  \\ inter-edge  at stage $i$.}\\
\hline
\end{tabular}
\end{center}
\label{tb:notation}
\end{table}
 }

\subsection{Stage $1:$ Random Failure of Nodes in Network $A$}
\label{sec:stage1}

Following the failures of a fraction $1-p$ of randomly selected
nodes in network $A$, the remaining network $\bar A_1$ has size
$pN$; since we eventually let $N$ grow large, $pN$ can be
approximated as an integer. As in \cite{Buldyrev,
newman2002spread,newman2001random,ShaoBuldyrevHavlinStanley}, we
use the technique of generating functions to quantify the fraction
of the functioning giant component $A_1 \subset \bar A_1$.
Specifically, let the function $P_A(p)$ determine the fraction of
the giant component in a random subgraph that occupies a fraction
$p$ of the nodes in network $A$ (the exact calculation of $P_A(p)$
will be elaborated later). It follows that the functioning giant
component has size
\begin{equation}
|A_1|=p P_{A}(p)N:=p_{A1}N. \label{eq:A_1'_giant}
\end{equation}

As shall become apparent soon, at the end of each stage it is
necessary to determine not only the size of the functioning giant
component, but also the specific inter-edge distribution over the
functioning nodes; i.e., the numbers of functioning nodes having
particular numbers of inter-edges. {\em Indeed, this is what makes
the analysis of the regular allocation model more complicated than
the models considered in
\cite{{Buldyrev},{ParshaniBuldyrevHavlin},{ShaoBuldyrevHavlinStanley}}.}
Here, at the end of Stage $1$, each node in $A_1$ still has $k$
inter-edges from network $B$ since network $B$ has not changed
yet.

\subsection{Stage $2:$ Impact of Random Node Failures in Network $A$ on Network $B$}

As the functioning part of network $A$ fragments from $A$ to $A_1$
(in Stage $1$), some of the inter-edges that were supporting
$B$-nodes would be removed. Observe that the probability of
removal can be approximated by $1-|A_1|/|A|= 1 - p_{A1}$ for each
inter-edge. With this perspective, a $B$-node loses $k-j$ of its
inter-edges with probability ${k \choose j}
p_{A1}^{j}(1-p_{A1})^{k-j}$. Moreover, it stops functioning with
probability $(1-p_{A1})^k$ due to losing all $k$ of its
inter-edges. As a result, with $\bar B_2$ denoting the set of
nodes in $B$ that retain at least one inter-edge, we have
 \begin{equation}
 |\bar B_2|= \left(1- (1-p_{A1})^k\right) N = p_{B2}'N, \label{B2'}
\end{equation}
where $ p_{B2}' = 1- (1-p_{A1})^k$. Also, the distribution of
inter-edges over the nodes in $\bar B_{2}$  is given by
\begin{equation} |\bar B_{2}|_{j} = {k \choose j} p_{A1}^{j}(1-p_{A1})^{k-j}
N, \quad j=1,2 \ldots, k, \label{eq:B_2_bar_dist}
\end{equation}
with $|\bar B_{2}|_{j}$ denoting the number of nodes in $\bar
B_{2}$ that have $j$ inter-edges.

As in Stage $1$, the size of  the functioning giant component $B_2
 \subset \bar B_{2}$ can be predicted by
\begin{equation}
|B_2| = p_{B2}' P_{B}(p_{B2}')N = p_{B2} N,
 \label{B2}
\end{equation}
where $P_B (\cdot) $ is defined analogously to the definition of
$P_A (\cdot)$ given in Section \ref{sec:stage1}. Obviously, each
node in $\bar B_2$ can survive as a functioning node in $B_2$ with
probability $P_B(p_{B2}')$. Thus, for each $j=1,2,\ldots, k$, the
number of nodes in $B_2$ that have $j$ inter-edges is given (in
view of (\ref{eq:B_2_bar_dist})) by
\begin{equation}
|B_{2}|_{j} = P_B(p_{B2}') {k \choose j}
p_{A1}^{j}(1-p_{A1})^{k-j} N. \label{eq:B2_dist}
\end{equation}

\subsection{Stage $3:$ Further A-Nodes Failures due to B-Node Failures}

Due to the fragmentation of the functional part of network $B$
from $\bar B_2$ to $B_2$ (not $B$ to $B_2$), some of the nodes in
$A_1$ may now lose all their inter-edges and stop functioning. To
compute the probability of this event, first observe that each
inter-edge from $\bar B_2$ to $A_1$ will be removed with an
approximate probability of $ 1-|B_2|/|\bar B_2| = 1 - P_B (p'_{B2}
)$. Hence the probability that a node in $A_1$ will lose all of
its inter-edges is given by $(1-P_B(p_{B2}'))^k$. It also follows
that the size of the network $\bar A_3 \subset A_1$ comprised of
the nodes that did not lose all their inter-connections is given
via
\begin{equation}
|\bar A_3|= p_{A1} \left(1-(1-P_B(p_{B2}'))^k\right) N.
\label{eq:A3'}
\end{equation}
In other words, in passing from $A_1$ to $\bar A_3$, a fraction
$1-|\bar A_3|/|A_1| = (1 - P_B (p'_{B2} ))^k$ of the nodes have
failed. As previously, the next step is to compute the size of the
functioning giant component $A_3 \subset \bar A_3$. However, this
a challenging task as noted in \cite{Buldyrev}. Instead, we view
the joint effect of the node failures in Stage $1$ and Stage $3$
as equivalent (in terms of the size of the resulting functional
giant component; i.e., $|A_3|$) to the effect of an initial random
attack that targets an appropriate fraction (to be determined
later) of the nodes. Intuitively, the node failures in $A_1$ at
Stage 3 (i.e., the removal of a fraction $(1 - P_B (p'_{B2} ))^k$
of nodes from $A_1$) have the same effect as taking out the same
portion from $\bar A_1$ \cite{Buldyrev}. In other words, it is
equivalent to the removal of a fraction $p(1 - P_B (p'_{B2} ))^k$
of the nodes from $A$. Recalling also that a fraction $1-p$ of the
nodes in network $A$ failed as a result of the initial attack at
Stage 1, we find that the fragmentation of $A$ to $\bar A_3$ can
as well be modeled (with respect to the size of $A_3$) by an
initial attack targeting a fraction
\[
1-p+ p\left(1-P_B(p_{B2}')\right)^k= 1- p
\left(1-(1-P_B(p_{B2}'))^k\right)
\]
of the nodes. It is now a standard step to conclude that, with
$p_{A3}'= p \left(1-(1-P_B(p_{B2}'))^k\right)$, the size of the
functioning giant component $A_3$ is given by
\begin{equation}
|A_3|=p_{A3}' P_A(p_{A3}') N = p_{A3} N . \label{eq:A3}
\end{equation}

\subsection{Stage $4:$ Further Fragmentation of Network B}

Due to the network fragmentation from $\bar A_3$ to $A_3$ in Stage
$3$, each inter-edge supporting a $B_2$-node will be disconnected
with probability that equals the proportion nodes in $\bar A_3$
that did not survive to $A_3$; i.e.,  $1-|A_3|/|\bar A_3|=1-
P_A(p'_{A3} )/P_A (p)  $ by (\ref{eq:A3'}) and (\ref{eq:A3}).
Consequently, a node in $B_2$ with $j$ inter-edges will stop
functioning with probability $\left( 1 -P_A (p'_{A3})/P_A (p)
\right) ^j$. Recalling also the inter-edge distribution
(\ref{eq:B2_dist}), the fraction $L$ of node failures in $B_2$ is
given by
\begin{eqnarray}\nonumber
\lefteqn{L} &&
\\ \nonumber
 &=& \frac{1}{N}\sum\limits_{j = 1}^k {\left| {B_2 } \right|_j
\left( {1 - \frac{{P_A (p'_{A3} )}}{{P_A (p)}}} \right)^j }
\\ \nonumber
 &=& P_B (p'_{B2} )   \sum\limits_{j = 1}^k {k \choose j} p_{A1}^j  (1 - p_{A1} )^{k - j}
 \left( {1 - \frac{{P_A (p'_{A3} )}}{{P_A (p)}}} \right)^j
\\ \nonumber
& =& P_B (p'_{B2} )\left( {\left( {1 - p_{A1} \frac{{P_A (p'_{A3}
)}}{{P_A (p)}} } \right)^k  - (1 - p_{A1} )^k } \right)
\\ \nonumber
&=& P_B (p'_{B2} )\left( {\left( {1 - pP_A (p'_{A3} )} \right)^k -
(1 - p_{A1} )^k } \right).
\end{eqnarray}
Since $|\bar B_4| = |B_2|- L N$, it follows that
\begin{equation} |\bar B_4| =
P_B(p_{B2}') \left(1-\left(1-p P_A(p_{A3}')\right)^{k}\right) N.
\label{eq:B_4}
\end{equation}

In order to compute the size of the functioning giant component
$B_4 \subset \bar B_4$, we proceed as in Stage $3$. Specifically,
we view the joint effect of node removals in Stage $2$ and Stage
$4$ as equivalent to that of an initial random attack which
targets an appropriate fraction of the nodes. To determine this
fraction, first observe that the failures in Stage $3$ have
triggered further node failures in $B_2$ resulting a fraction
\begin{equation} \label{eq:b2f}
1 - {{|\bar B_4 |} \mathord{\left/
 {\vphantom {{|\bar B_4 |} {|B_2 |}}} \right.
 \kern-\nulldelimiterspace} {|B_2 |}}
=1 - {{\left( {1 - (1 - pP_A (p'_{A3} ))^k } \right)}
\mathord{\left/
 {\vphantom {{\left( {1 - (1 - pP_A (p'_{A3} ))^k } \right)} {p'_{B2} }}} \right.
 \kern-\nulldelimiterspace} {p'_{B2} }}
\end{equation}
of the nodes' failure. Next, note that the effect of these
failures on $|B_4|$ is equivalent to that of taking out the same
fraction of nodes from $\bar B_2$ \cite{Buldyrev}. Moreover, it
has the same effect as taking out a fraction $ p'_{B2} \left\{ {1
- {{\left( {1 - (1 - pP_A (p'_{A3} ))^k } \right)} \mathord{\left/
 {\vphantom {{\left( {1 - (1 - pP_A (p'_{A3} ))^k } \right)} {p'_{B2} }}} \right.
 \kern-\nulldelimiterspace} {p'_{B2} }}} \right\}$
of the nodes in $B$. Now, recalling that a fraction $1-p_{B2}'$ of
nodes in $B$ have failed in Stage $2$, we conclude that the joint
effect of cascading failures in Stage $2$ and Stage $4$ (on
$|B_4|$) is identical to that of an initial random attack which
targets a fraction
\begin{eqnarray}
\lefteqn{
 1-p_{B2}' + p_{B2}'\left(1-\frac{1-\left(1-p
P_A(p_{A3}')\right)^{k}}{p_{B2}'}\right)} &&
\nonumber \\
&=& \left(1-p P_A(p_{A3}')\right)^{k}\hspace{3cm} \nonumber
\end{eqnarray}
of nodes. As previously, with $p_{B4}' = 1- \left(1-p
P_A(p_{A3}')\right)^{k}$ we conclude that the size of the
functioning giant component $B_4$ is given by
$|B_4|=p_{B4}'P_B(p_{B4}')N=p_{B4} N$.

\subsection{Cascading Dynamics of Node Failures}
\label{sec:recursive1} As mentioned earlier, the main goal of this
section is to characterize the size of the functional giant
components in steady state. Indeed, along the lines outlined
above, one can obtain the sizes of all functioning giant
components $A_1\supset A_3 \supset \ldots \supset A_{2m+1}$ and
$B_2 \supset B_4 \supset \ldots \supset B_{2m}$ for any integer
$m$. However, it is easy to observe the pattern in the expressions
obtained so far and conclude that with $p_{A1}'=p$ the size of all
giant components are given by the recursive relations
\begin{eqnarray}  \label{eq:recursive_sys_1}
 \begin{array}{l}
 p_{Ai}=p_{Ai}'P_A(p_{Ai}'), \\
 p_{Ai}'=p\left(1-\left(1-P_B(p_{Bi-1}')\right)^k\right),
  \end{array}
{i=3,5,7\ldots}
\end{eqnarray}
and
\begin{equation}
 \label{eq:recursive_sys_1b}
 \begin{array}{l}
 p_{Bi}=p_{Bi}'P_B(p_{Bi}'), \\
 p_{Bi}'=1-\left(1-pP_A(p_{Ai-1}')\right)^k, \\
 \end{array}
 {i=2,4,6,\ldots.}
\normalsize
\end{equation}
This recursive process stops at an \lq\lq equilibrium point" where
we have $p'_{B2m-2}=p'_{B2m}$ and $p'_{A2m-1}=p'_{A2m+1}$ so that
neither network A nor network B fragments further. Setting
$x=p'_{A2m+1}$ and $y=p'_{B2m}$, this yields the transcendental
equations
\begin{equation}
x=p \left(1-\left(1-P_B(y)\right)^k\right) \quad
y=1-\left(1-pP_A(x)\right)^k. \label{eq:system1}
\end{equation}

The analysis carried out up to this point is valid for {\em all}
networks, irrespective of their intra-structures. In principle,
for specific intra-structures of networks $A$ and $B$ (which
determine the functions $P_A$ and $P_B$, respectively), the system
(\ref{eq:system1}) of equations can be solved for given $p$ and
$k$. The steady-state fractions of nodes in the giant components
can then be computed by using the relations $\lim_{i \to
\infty}p_{Ai} := P_{A_{\infty}}=xP_A(x)$ and $\lim_{i \to
\infty}p_{Bi} :=P_{B_{\infty}}=yP_B(y)$. Indeed, in Section
\ref{sec:Numerical}, we consider a special case where both
networks $A$ and $B$ are Erd\H{o}s-R\'{e}nyi (ER) graphs
\cite{Bollobas} and give solutions of the system
(\ref{eq:system1}) for several values of $p$ and $k$.

\section{Optimality of Regular Allocation Strategy}
\label{sec:proof of regular}

In this section, we show analytically that the regular allocation
strategy always yields stronger robustness than other strategies
and thus it is optimal in the absence of intra-topology
information. In the following, we refer to the system that uses
the regular allocation strategy as System $1$. Specifically, we
consider two networks $A$ and $B$ where each node is uniformly
supported by $k$ bi-directional inter-edges. For convenience, we
denote the fractions in the recursive relations
(\ref{eq:recursive_sys_1})-(\ref{eq:recursive_sys_1b}) as
$p'_{Ai}(p;k)$ and $p'_{Bi}(p;k)$, where $1-p$ is the initial
fraction of failed nodes in network $A$. Also, we let
$P_{A^1_{\infty}}(p ; k)$ and $P_{B^1_{\infty}}(p ; k)$ be the
steady-state fractions of functional giant components of the two
networks, respectively. Finally, we use $p_{c_1}(k)$ to denote the
critical threshold associated with System $1$.

In what follows, we first investigate the dynamics of cascading
failures in the auxiliary System $2$, where bi-directional
inter-edges are distributed {\em randomly} amongst nodes. The
analysis is carried out under a {\em generic} inter-degree
distribution so that all possible (bi-directional) inter-link
allocation strategies are covered. By making use of the convexity
property and Jensen's inequality, we show that for a fixed mean
inter-degree, System $2$ achieves the highest robustness against
random attacks when its inter-degree distribution degenerates,
i.e.; when all nodes have exactly the same number of inter-edges
so that System $2$ is equivalent to System $1$. Therefore, we
conclude that regular allocation yields the strongest robustness
amongst all possible (bi-directional) inter-link allocation
strategies. Next, we show that systems with bi-directional
inter-edges can better combat the cascading failures compared to
the systems with unidirectional inter-edges
\cite{ShaoBuldyrevHavlinStanley}. Together, these results prove
the optimality of the inter-link allocation strategy in System
$1$; i.e., regular allocation of bi-directional inter-edges.

\subsection{Analysis of  Random Allocation Strategy}
\label{subsec:comparing_random}

We now introduce the auxiliary System $2$. Consider two arbitrary
networks $A$ and $B$, each with $N$ nodes, and a discrete
probability distribution $F:\mathbb{N}\to[0,1]$, such that
\begin{equation}\label{eq:F_distribution}
F(j)=\alpha_j, \quad j=0,1,\ldots ,
\end{equation}
with $ \sum_{j=0}^{\infty}\alpha_j=1.$

To allocate the interdependency links, we first partition each
network randomly into subgraphs with sizes $\alpha_0 N, \alpha_1
N, \alpha_2 N, \ldots$. \footnote{For $N$ large enough, each of
these subgraph sizes can be well approximated by an integer.} By
doing so,
 we can obtain subgraphs $\{S_{A_{\alpha_0}},S_{A_{\alpha_1}},
S_{A_{\alpha_{2}}},\ldots\}$ and
$\{S_{B_{\alpha_0}},S_{B_{\alpha_1}},S_{B_{\alpha_{2}}},\ldots\}$,
such that
\[
|S_{A_{\alpha_j}}|=|S_{B_{\alpha_j}}|=\alpha_j N, \quad
j=0,1,\ldots.
\]

Then, for each $j=0,1,\ldots$, assume that each node in the
subgraphs $S_{A_{\alpha_j}}$ and $S_{B_{\alpha_j}}$ is assigned
$j$ bi-directional inter-edges. This ensures that the inter-degree
of each node is a {\em random} variable drawn from the
distribution $F$; i.e., an arbitrary node will have $j$
inter-edges with probability $\alpha_j$, for each $j=0,1,\ldots$.
It is worth noting that the inter-degrees of the nodes are {\em
not} mutually independent since the total number of inter-edges is
fixed at $E = \sum\limits_{}^{} {\alpha _j j N}$ for both
networks. 

We have a few more words on the possible implementation of the
above random allocation strategy. Observe that each bi-directional
edge can be treated equivalently as two unidirectional edges. In
this way, there are a total of $2E$ unidirectional inter-edges in
the system, where $E$ edges are going outward from network $A$ and
the other $E$ edges are going outward from network $B$.  We
randomly match each unidirectional edge going outward from $A$ to
a unique edge going outward from $B$ and combine them into a
single bi-directional edge. To this end, let the edges going
outward from $A$ and $B$ be separately labeled as
$\boldsymbol{e}=\{e_1,\ldots,e_E\}$ and
$\boldsymbol{e'}=\{e_1',\ldots,e_E'\}$, respectively. Next, use
the Knuth shuffle algorithm \cite{knuth1981art} to obtain random
permutations $\boldsymbol{\bar{e}}=\{\bar{e}_1,\ldots,\bar{e}_E\}$
and $\boldsymbol{\bar{e}'}=\{\bar{e}'_1,\ldots,\bar{e}'_E\}$ of
the vectors $\boldsymbol{e}$ and $\boldsymbol{e'}$, respectively.
Finally, for each $i=1,\ldots, E$, match the unidirectional
inter-edges $\bar{e}_j$ and $\bar{e'}_j$  to obtain $E$
bi-directional inter-edges.

We now analyze the dynamics of cascading failures in System $2$
using an iterative approach similar to that in Section
\ref{sec:analysis}.  For brevity, we skip most of the details and
give only an outline of the arguments that lead to the sizes of
functional giant components. The main difference from the analysis
of Section \ref{sec:analysis} is that the fractions of nodes in
$A$ and $B$ retaining at least one inter-edge, i.e., the fractions
$\bar A_i$ and $\bar B_i$, need to be calculated differently from
(\ref{eq:A3'}) and (\ref{eq:B_4}) due to the random inter-degree
of each node.

Owing to the fragmentation from $\bar B_{i-1}$ to $B_{i-1}$, each
inter-edge supporting $A$ could be disconnected with probability
$1-|B_{i-1}|/|\bar B_{i-1}|$, triggering further failures in
network $A$ at step $i$. With this insight, the aggregate effect
of the
 failures in $B$ up to stage $i$ can be treated equivalently
(with respect to the size of $A_i$) as removing each inter-edge
supporting $A$ with probability $1 - u_i$.  According to Section
\ref{sec:analysis}, $u_i$ can be derived as follows:
\begin{equation}
u_i  = \prod\limits_{\ell = 1}^{(i - 1)/2} {\frac{{\left|
{B_{2\ell} } \right|}}{{\left| {\bar B_{2\ell} } \right|}}}  = P_B
(p'_{Bi - 1} )~~~~i=3,5,7...,
\end{equation}
Similarly, the aggregate effect of node failures in $A$ before
step $i$ can be viewed as equivalent to removing each inter-edge
supporting $B$ with probability $1 - v_i$ (with respect to the
size of $B_i$) such that
\begin{equation}
v_i  = \frac{{\left| {A_1 } \right|}}{{\left| A
\right|}}\prod\limits_{\ell = 1}^{i/2 - 1} {\frac{{\left|
{A_{2\ell + 1} } \right|}}{{\left| {\bar A_{2\ell + 1} }
\right|}}} = pP_A (p'_{Ai - 1} ) ~~~~i=2,4,6....
\end{equation}

In System $2$, each node is supported by $j$ inter-edges with
probability $\alpha_j$. In view of this, at  step $i$, a node in
network $A$ would retain at least one inter-edge with probability
$1 - \sum\limits_{j=0}^{\infty} {\alpha _j (1 - u_i )^{j } }$.
Recalling also that a fraction $1-p$ of the nodes had already
failed before the onset of the cascading failure, the equivalent
remaining fraction of network $A$ at stage $i$ is given by:
\begin{eqnarray} \nonumber
 p'_{Ai}  &=& p(1 - \sum\limits_{j=0}^{\infty} {\alpha _j (1 - u_i )^{j } } )
 \\  \nonumber
  &=& p(1 - \sum\limits_{j=0}^{\infty} {\alpha _j (1 - P_B (p'_{Bi - 1} ))^{j } }
  ).
\end{eqnarray}
Similarly, the equivalent remaining fraction of network $B$ turns
out to be
\begin{eqnarray}  \nonumber
 p'_{Bi}  &=& 1 - \sum\limits_{j=0}^{\infty} {\alpha _j (1 - v_i )^{j } }
 \\  \nonumber
  &=& 1 - \sum\limits_{j=0}^{\infty} {\alpha _j (1 - pP_A (p'_{Ai - 1} ))^{j }
  }.
\end{eqnarray}

Hence, the fractional sizes of the giant components at each stage
are given (with $p'_{A1}=p$) by
\begin{equation}  \label{eq:recursive_sys_21}
 \begin{array}{l}
 p_{Ai}=p_{Ai}'P_A(p_{Ai}'),\\
 p'_{Ai}=p(1 - \sum\limits_{j=0}^{\infty} {\alpha _j (1 - P_B (p'_{Bi - 1} ))^j }), \\
  \end{array}
\end{equation}
for $i=3,5,7\ldots,$ and by
\begin{equation}    \label{eq:recursive_sys_22}
 \begin{array}{l}
p_{Bi}=p_{Bi}'P_B(p_{Bi}'),\\
p_{Bi}'=1 - \sum\limits_{j=0}^{\infty} {\alpha _j (1 - pP_A
(p'_{Ai - 1} ))^j
  }, \\
  \end{array}  \\
\end{equation}
for $i=2,4,6, \ldots$. We next show that System $1$ is always more
robust than System $2$ against random attacks by comparing the
recursive relations
(\ref{eq:recursive_sys_1})-(\ref{eq:recursive_sys_1b}) and
(\ref{eq:recursive_sys_21})-(\ref{eq:recursive_sys_22}).

\subsection{Regular Allocation versus Random Allocation}
\label{subsec:reg_vs_rand}

We now compare Systems $1$ and $2$ in terms of their robustness
against random attacks. For convenience, we use a vector
$\boldsymbol{\alpha}=(\alpha_0, \alpha_1, \ldots)$ to characterize
the inter-degree distribution $F$, where $F(j)=\alpha_j$. Next, we
denote the fractions in the recursive relations
(\ref{eq:recursive_sys_21})-(\ref{eq:recursive_sys_22}) as
$p_{Ai}(p;\boldsymbol{\alpha}), p'_{Ai}(p;\boldsymbol{\alpha})$
and $p_{Bi}(p;\boldsymbol{\alpha}),
p'_{Bi}(p;\boldsymbol{\alpha})$. Also, we let $P_{A_{\infty}^2}(p
; \boldsymbol{\alpha})$ and $P_{B_{\infty}^2}(p ;
\boldsymbol{\alpha})$ be the respective steady-state fractions of
the functional giant components in the two networks where $1-p$ is
the
 fraction  of initially failed nodes in network $A$.  In other
words, we set $\lim_{i \to \infty}p_{Ai} (p ; \boldsymbol{\alpha})
:= P_{A_{\infty}^2}(p ; \boldsymbol{\alpha})$ and $\lim_{i \to
\infty}p_{Bi}(p ; \boldsymbol{\alpha}):=P_{B_{\infty}^2}(p ;
\boldsymbol{\alpha})$. Finally, we denote the critical threshold
associated with System $2$ by $p_{c_2}(\boldsymbol{\alpha})$.

Assume that network $A$ (respectively network $B$) of Systems $1$
and $2$ have the same size $N$ and   the same
 intra-degree distribution such that the functions $P_A$ (respectively $P_B$) are identical for
both systems. The next result shows that if the two systems are
\lq\lq matched" through their mean inter-degrees, i.e., if $k =
\sum\limits_{j = 0}^\infty {\alpha _j j}$, System $1$ always
yields stronger robustness than System $2$ against random node
failures.
\begin{thm}
{\sl Under the condition
\begin{equation}
k = \sum\limits_{j = 0}^\infty {\alpha _j j},
\label{eq:matching_condition}
\end{equation}
we have
\begin{equation}
\begin{array}{l}
  P_{A^1_{\infty}}(p ; k) \geq P_{A^2_{\infty}}(p ; \boldsymbol{\alpha}), \\
  P_{B^1_{\infty}}(p ; k) \geq P_{B^2_{\infty}}(p ;
  \boldsymbol{\alpha});
 \end{array}
\label{eq:Giants_Sys1_geq_Sys2}
\end{equation}
and furthermore
\begin{equation}
p_{c_1}(k) \leq p_{c_2}(\boldsymbol{\alpha}).
\label{eq:p_c_sys1_leq_sys2}
\end{equation}
\label{thm:sys_1_robust_sys2} }
\end{thm}

\bproof Since  $P_A$ and $P_B$ are monotonically increasing
functions \cite{newman2001random}, a sufficient condition ensuring
 (\ref{eq:Giants_Sys1_geq_Sys2}) will
hold is
\begin{equation}
\begin{array}{l}
p'_{Ai}(p;k) \geq p'_{Ai}(p;\boldsymbol{\alpha}), \quad i=3,5,7 \ldots,\\
p'_{Bi}(p;k) \geq p'_{Bi}(p;\boldsymbol{\alpha}), \quad i=2,4,6
\ldots,
 \end{array}
\label{eq:to_show_prop1}
\end{equation}
where $p'_{Ai}(p;k)$, $p'_{Bi}(p;k)$, and
$p'_{Ai}(p;\boldsymbol{\alpha})$, $p'_{Bi}(p;\boldsymbol{\alpha})$
denote the fractions in the recursive relations
(\ref{eq:recursive_sys_1})-(\ref{eq:recursive_sys_1b}), and
(\ref{eq:recursive_sys_21})-(\ref{eq:recursive_sys_22}),
respectively. We establish (\ref{eq:to_show_prop1}) by induction.
First observe that $p'_{A1}(p;k)=p'_{A1}(p;\boldsymbol{\alpha})=p$
and the inequality (\ref{eq:to_show_prop1}) is satisfied for
$i=1$. In view of
(\ref{eq:recursive_sys_1})-(\ref{eq:recursive_sys_1b}) and
(\ref{eq:recursive_sys_21})-(\ref{eq:recursive_sys_22}), condition
(\ref{eq:to_show_prop1}) for $i=2$ will be satisfied if
\begin{equation}\nonumber
 \left( {1 - pP_A (p'_{A1}(p;k)}
\right)^k \le \sum\limits_{j = 0}^{\infty} {\alpha _j \left( {1 -
pP_A (p'_{A1}(p;\boldsymbol{\alpha}) )} \right)^j },
\end{equation}
or equivalently
\begin{equation}
 \left( {1 - pP_A (p)}
\right)^k \le \sum\limits_{j = 0}^{\infty} {\alpha _j \left( {1 -
pP_A (p)} \right)^j }. \label{eq:ind_1}
\end{equation}
Under (\ref{eq:matching_condition}), the convexity of $(1 - pP_A
(p))^x$ implies (\ref{eq:ind_1}) by Jensen's inequality. Hence, we
get that $p'_{B2}(p;k) \geq p'_{B2}(p;\boldsymbol{\alpha})$ and
the base step is completed.

Suppose that the condition (\ref{eq:to_show_prop1}) is satisfied
for each $i=1,2,\ldots, 2m-1, 2m$. We need to show that
(\ref{eq:to_show_prop1}) holds also for $i=2m+1$ and $i=2m+2$. For
$i=2m+1$, the first inequality will be satisfied if it holds that

\begin{equation}\nonumber
\left( {1 - P_B (p'_{B2m}(p;k) )} \right)^k  \le \sum\limits_{j =
0}^{\infty} {\alpha _j \left( {1 - P_B (p'_{B2m
}(p;\boldsymbol{\alpha}) )} \right)^j }
\end{equation}
By the induction hypothesis, we have $P_B (p'_{B2m}(p;k) ) \geq
P_B (p'_{B2m }(p;\boldsymbol{\alpha}) )$ since $p'_{B2m}(p;k) \geq
p'_{B2m }(p;\boldsymbol{\alpha})$. As a result, the above
inequality is satisfied if
\begin{equation}\label{eq:ind_2}
\left( {1 - u} \right)^k  \le \sum\limits_{j = 0}^{\infty} {\alpha
_j \left( {1 - u} \right)^j }
\end{equation}
with $u=P_B (p'_{B2m }(p;\boldsymbol{\alpha}) )$. As before, under
(\ref{eq:matching_condition}), (\ref{eq:ind_2}) is ensured by the
convexity of $(1-u)^x$ in view of Jensen's inequality. The
condition $p'_{A2m+1}(p;k) \geq p'_{A2m+1}(p;\boldsymbol{\alpha})$
is now established.

Now let $i=2m+2$. The desired condition $p'_{B2m+2}(p;k) \geq
p'_{B2m+2}(p;\boldsymbol{\alpha})$ will be established if
\begin{eqnarray}
\lefteqn{\left(1 - pP_A (p'_{A2m+1}(p;k)) \right)^k} &&
\nonumber \\
&\leq& \sum\limits_{j = 0}^{\infty} {\alpha _j \left( {1 - pP_A
(p'_{A2m+1}(p;\boldsymbol{\alpha}) )} \right)^j }, \nonumber
\end{eqnarray}
or equivalently
\begin{equation}
\left( {1 - v} \right)^k \le \sum\limits_{j = 0}^{\infty} {\alpha
_j \left( {1 - v} \right)^j }, \label{eq:ind_3}
\end{equation}
where we set $v=pP_A (p'_{A2m+1}(p;\boldsymbol{\alpha}) )$. The
last step follows from the previously obtained fact that
$p'_{A2m+1}(p;k) \geq p'_{A2m+1}(p;\boldsymbol{\alpha})$. Once
more, (\ref{eq:ind_3}) follows by the convexity of $(1-v)^x$ and
Jensen's inequality. This establishes the induction step and the
desired conclusion (\ref{eq:Giants_Sys1_geq_Sys2}) is obtained.

We next prove the inequality $p_{c_1}(k) \leq
p_{c_2}(\boldsymbol{\alpha})$ by way of contradiction. Assume
towards a contradiction that $p_{c_2}(\boldsymbol\alpha) <
p_{c_1}(k)$ and fix $p$ such that $p_{c_2}(\boldsymbol{\alpha}) <
p < p_{c_1}(k)$. Then, let a fraction $1-p$ of the nodes randomly
fail in network $A$ of both systems. Since $p$ is less than
$p_{c_1}$, the node failures will eventually lead to complete
fragmentation of the two networks in System $1$; i.e., we get
$P_{A^1_{\infty}}(p ; k) = P_{B^1_{\infty}}(p ; k)=0$. On the
other hand, the fact that $p$ is larger than the critical
threshold $p_{c_2}$ ensures $P_{A^2_{\infty}}(p ;
\boldsymbol{\alpha})
>0$ and $P_{B^2_{\infty}}(p ; \boldsymbol{\alpha})>0$ by definition. This
clearly contradicts (\ref{eq:Giants_Sys1_geq_Sys2}) and therefore
it is always the case that $p_{c_1}(k) \leq
p_{c_2}(\boldsymbol{\alpha})$ under (\ref{eq:matching_condition}).
\eproof

We have now established that the regular allocation of
bi-directional inter-edges always yields  stronger robustness than
any possible random allocation strategy that uses bi-directional
links. In the following section, we show that using bidirectional
inter-edges leads to a smaller critical threshold and better
robustness than using unidirectional inter-edges.

\subsection{Bi-directional Inter-Edges versus Unidirectional Inter-Edges}
\label{subsec:bi_uni}

We now compare the robustness of System $2$ with that of the model
considered in \cite{ShaoBuldyrevHavlinStanley}, hereafter referred
to as System $3$. As mentioned earlier, the model considered in
\cite{ShaoBuldyrevHavlinStanley} is based on the random allocation
of {\em unidirectional} inter-edges and can be described as
follows. As with System $2$, consider two arbitrary networks $A$
and $B$, each with $N$ nodes, and a discrete probability
distribution $F:\mathbb{N}\to[0,1]$ such that
(\ref{eq:F_distribution}) holds. Assume that each node is
associated with a random number of supporting nodes from the other
network, and that this random number is distributed according to
$F$. In other words, for each $j=0,1,\ldots$, a node has $j$ {\em
inward} inter-edges with probability $\alpha_j$. The supporting
node for each of these inward edges is selected randomly amongst
all nodes of the other network ensuring that the number of {\em
outward} inter-edges follows a {\em binomial} distribution for all
nodes.

System $3$ was studied in \cite{ShaoBuldyrevHavlinStanley} using
similar methods to those of Section \ref{sec:analysis} and Section
\ref{subsec:comparing_random}. This time, after an initial failure
of a fraction $1-p$ of the nodes in network $A$, the recursive
relations for the fractions of giant components at each stage
turns \cite{ShaoBuldyrevHavlinStanley} out to be (with
$p'_{A1}=p$)
\begin{equation}  \label{eq:recursive_sys_31}
 \begin{array}{l}
 p_{Ai} =p_{Ai}'P_A(p_{Ai}'),\\
p'_{Ai}=p(1 - \sum\limits_{j = 0}^{\infty} {\alpha_j (1 - p'_{Bi - 1} P_B (p'_{Bi - 1} ))^j }), \\
  \end{array}
\end{equation}
for $i=3,5,7\ldots,$ and
\begin{equation}    \label{eq:recursive_sys_32}
 \begin{array}{l}
p_{Bi}=p_{Bi}'P_B(p_{Bi}'),\\
p_{Bi}'=1 - \sum\limits_{j = 0}^{\infty} {\alpha_j (1 - p'_{Ai -
1} P_A (p'_{Ai - 1} ))^j
  }, \\
  \end{array}  \\
\end{equation}
for  $i=2,4,6, \ldots$.

Next, we compare System $2$ and System $3$ using the recursive
relations (\ref{eq:recursive_sys_21})-(\ref{eq:recursive_sys_22})
and (\ref{eq:recursive_sys_31})-(\ref{eq:recursive_sys_32}). In
doing so, we use the same notation to define the fractions in the
recursive relations
(\ref{eq:recursive_sys_21})-(\ref{eq:recursive_sys_22}) as used in
Section \ref{subsec:reg_vs_rand}, while the fractions in
(\ref{eq:recursive_sys_31})-(\ref{eq:recursive_sys_32}) will be
denoted by $p^{3}_{Ai}(p;\boldsymbol{\alpha}),
{p'}^{3}_{Ai}(p;\boldsymbol{\alpha})$ and
$p^3_{Bi}(p;\boldsymbol{\alpha}),
{p'}^3_{Bi}(p;\boldsymbol{\alpha})$. We let $P_{A_{\infty}^3}(p ;
\boldsymbol{\alpha})$ and $P_{B_{\infty}^3}(p ;
\boldsymbol{\alpha})$ be the steady-state fractions of functional
giant components in System $3$ if a fraction $1-p$ of the nodes
initially  fail in network $A$. In other words, we set $\lim_{i
\to \infty}p^{3}_{Ai} (p ; \boldsymbol{\alpha}) :=
P_{A_{\infty}^3}(p ; \boldsymbol{\alpha})$ and $\lim_{i \to
\infty}p^{3}_{Bi}(p ; \boldsymbol{\alpha}):=P_{B_{\infty}^3}(p ;
\boldsymbol{\alpha})$. Finally, we denote by
$p_{c_3}(\boldsymbol{\alpha})$ the critical threshold for System
$3$.

The next result shows that System $2$ is always more robust than
System $3$ against random node failures.

\begin{thm}
{\sl We have that
\begin{equation}
\begin{array}{l}
  P_{A^2_{\infty}}(p ; \boldsymbol{\alpha}) \geq P_{A^3_{\infty}}(p ; \boldsymbol{\alpha}), \\
  P_{B^2_{\infty}}(p ; \boldsymbol{\alpha}) \geq P_{B^3_{\infty}}(p ; \boldsymbol{\alpha}), \\
 \end{array}
\label{eq:Giants_Sys2_geq_Sys3}
\end{equation}
and furthermore,
\begin{equation}
p_{c_2}(\boldsymbol{\alpha}) \leq p_{c_3}(\boldsymbol{\alpha}).
\label{eq:p_c_sys2_leq_sys3}
\end{equation}
\label{thm:sys_2_robust_sys3} }
\end{thm}

\bproof Since  $P_A(x)$ and $P_B(x)$ are monotonically increasing
\cite{newman2001random}, a sufficient condition ensuring
(\ref{eq:Giants_Sys2_geq_Sys3}) is given by
\begin{equation}
\begin{array}{l}
p'_{Ai}(p;\boldsymbol{\alpha}) \geq {p'}^{3}_{Ai}(p;\boldsymbol{\alpha}), \quad i=1,3,5 \ldots,\\
p'_{Bi}(p;\boldsymbol{\alpha}) \geq
{p'}^{3}_{Bi}(p;\boldsymbol{\alpha}), \quad i=2,4,6 \ldots.
 \end{array}
\label{eq:to_show_prop2}
\end{equation}
We establish (\ref{eq:to_show_prop2}) by  induction. First,
observe that for $i=1$, $p'_{A1}(p;\boldsymbol{\alpha}) =
{p'}^{3}_{A1}(p;\boldsymbol{\alpha}) = p$ and condition
(\ref{eq:to_show_prop2}) is satisfied. Next, for $i=2$, we see
from (\ref{eq:recursive_sys_22}) and (\ref{eq:recursive_sys_32})
that the inequality
\[
p'_{B2}(p;\boldsymbol{\alpha}) \geq
{p'}^{3}_{B2}(p;\boldsymbol{\alpha})
\]
will hold if
\begin{eqnarray}\label{eq:ind_4}
\lefteqn {\sum\limits_{j = 0}^{\infty} \alpha_j (1 - p P_A
(p'_{A1}(p;\boldsymbol{\alpha}) ))^j} &&
 \\ \nonumber
&\leq&  \sum\limits_{j = 0}^{\infty} \alpha_j (1 -
{p'}^{3}_{A1}(p;\boldsymbol{\alpha}) P_A
({p'}^{3}_{A1}(p;\boldsymbol{\alpha}) ))^j.
\end{eqnarray}
Since $p'_{A1}(p;\boldsymbol{\alpha}) =
{p'}^{3}_{A1}(p;\boldsymbol{\alpha}) = p$, it is immediate that
(\ref{eq:ind_4}) is satisfied with equality and this completes the
base step of the induction.

Suppose now that condition (\ref{eq:to_show_prop2}) is satisfied
for all $i=1,2,\ldots, 2m-1,2m$. We will establish
(\ref{eq:to_show_prop2}) for $i=2m+1$ and $i=2m+2$ as well.
Comparing (\ref{eq:recursive_sys_21}) and
(\ref{eq:recursive_sys_31}), it is easy to check that for
$i=2m+1$, (\ref{eq:to_show_prop2}) will hold if
\begin{eqnarray}\label{eq:ind_5}
\lefteqn {\sum\limits_{j = 0}^{\infty} \alpha_j (1 -P_B
(p'_{B2m}(p;\boldsymbol{\alpha}) ))^j} &&
\\
\nonumber &\leq&  \sum\limits_{j = 0}^{\infty} \alpha_j (1 -
{p'}^{3}_{B2m}(p;\boldsymbol{\alpha}) P_B
({p'}^{3}_{B2m}(p;\boldsymbol{\alpha}) ))^j.
\end{eqnarray}
By the induction hypothesis, (\ref{eq:to_show_prop2}) holds for
$i=2m$ so that $P_B ({p'}^{3}_{B2m}(p;\boldsymbol{\alpha})) \leq
P_B (p'_{B2m}(p;\boldsymbol{\alpha}) )$. It is now immediate that
(\ref{eq:ind_5}) holds, since we always have
${p'}^{3}_{B2m}(p;\boldsymbol{\alpha}) \leq 1$. This establishes
(\ref{eq:to_show_prop2}) for $i=2m+1$; i.e., that
\begin{equation}
{p'}^{3}_{A2m+1}(p;\boldsymbol{\alpha}) \leq
p'_{A2m+1}(p;\boldsymbol{\alpha}). \label{eq:ind_6}
\end{equation}

For $i=2m+2$, we see from (\ref{eq:recursive_sys_22}) and
(\ref{eq:recursive_sys_32}) that condition
(\ref{eq:to_show_prop2}) will be satisfied if
\begin{eqnarray}\label{eq:ind_7}
\lefteqn {\sum\limits_{j = 0}^{\infty} \alpha_j (1 - p P_A
(p'_{A2m+1}(p;\boldsymbol{\alpha}) ))^j} &&
\\ \nonumber
&\leq&  \sum\limits_{j = 0}^{\infty} \alpha_j (1 -
{p'}^{3}_{A2m+1}(p;\boldsymbol{\alpha}) P_A
({p'}^{3}_{A2m+1}(p;\boldsymbol{\alpha}) ))^j .
\end{eqnarray}
In view of (\ref{eq:ind_6}) and the fact that
${p'}^{3}_{A2m+1}(p;\boldsymbol{\alpha}) \leq p$, we immediately
obtain (\ref{eq:ind_7}) and the induction step is now completed.
This establishes condition (\ref{eq:to_show_prop2}) for all
$i=1,2,\ldots$, and we get (\ref{eq:Giants_Sys2_geq_Sys3}).

The fact that (\ref{eq:Giants_Sys2_geq_Sys3}) implies
(\ref{eq:p_c_sys2_leq_sys3}) can be shown by contradiction, as in
the proof of Theorem \ref{thm:sys_1_robust_sys2}. \eproof

Summarizing, it can be seen from Theorem
\ref{thm:sys_2_robust_sys3} that using bi-directional inter-edges
(System $2$) always yields stronger system robustness compared to
using unidirectional inter-edges (System $3$). This being valid
under an arbitrary distribution $\boldsymbol{\alpha}$ of
inter-edges, we conclude that regular allocation of bi-directional
inter-edges leads to the strongest robustness (amongst all
possible strategies) against random attacks as we recall Theorem
\ref{thm:sys_1_robust_sys2}.

\section{Numerical Results: The Erd\H{o}s-R\'{e}nyi Networks Case}
\label{sec:Numerical}

To get a  more concrete sense of the above analysis results, we
next look at some special cases of network models. In particular,
we assume both networks are Erd\H{o}s-R\'{e}nyi networks
\cite{Bollobas}, with mean intra-degrees  $a$ and $b$,
respectively. For this case, the functions $P_A(x)$ and $P_B(y)$
that determine the size of the giant components can be obtained
\cite{newman2001random} from
\begin{equation}
P_A(x)=1-f_A \quad \mbox{and} \quad P_B(y)=1-f_B, \label{eq:P_A_B}
\end{equation}
where $f_A$ and $f_B$ are the unique solutions of
\begin{equation}
 f_A=\exp\{ax(f_A-1)\} \quad \mbox{and} \quad f_B=\exp\{by(f_B-1)\}.
\label{eq:f_ab}
\end{equation}
In what follows, we derive numerical results for the steady-state
giant component sizes as well as critical $p_c$ values.
Specifically, we first study System $1$ by exploiting the
recursive relations
(\ref{eq:recursive_sys_1})-(\ref{eq:recursive_sys_1b}) using
(\ref{eq:P_A_B}) and (\ref{eq:f_ab}). Similarly, we derive
numerical results for System $2$ by using the recursive relations
(\ref{eq:recursive_sys_21})-(\ref{eq:recursive_sys_22}). For both
cases, we use extensive simulations to verify the validity of the
results obtained theoretically.

\subsection{Numerical Results for System $1$}
\label{subsec:num1}

Reporting (\ref{eq:P_A_B}) into (\ref{eq:system1}), we get
\begin{equation} x=p(1-f_B^k) \qquad y=1-
\left(1-p(1-f_A)\right)^k.
 \label{eq:system1b}
\end{equation}
It follows that the giant component fractions at steady state are
given by
\begin{equation}
\begin{array}{l}\small
  P_{A_{\infty}}=p(1-f_B^k)(1-f_A), \\
   P_{B_{\infty}}=\left(1-
\left(1-p(1-f_A)\right)^k\right)(1-f_B). \normalsize
\end{array}
  \label{eq:giant_comp}
\end{equation}
Next, substituting (\ref{eq:system1b}) into (\ref{eq:f_ab}) we
obtain
\begin{equation}
\begin{array}{ll}
f_A=\exp\{ap(1-f_B^k)(f_A-1)\}, & \\
f_B=\exp\{b \left(1- \left(1-p(1-f_A)\right)^k\right)(f_B-1)\}.
\end{array}
\label{eq:system3}
\end{equation}
We note that the system of equations (\ref{eq:system3})  always
has a trivial solution $f_A=f_B=1$, in which case the  functional
giant component has zero fraction for both networks. More
interesting cases arise for large values of $p$ when there exist
non-trivial solutions to (\ref{eq:system3}). In particular, we
focus on determining the critical threshold $p_c$; i.e., the {\em
minimum} $p$ that yields a non-trivial solution of the system.
Exploring this further, we see by elementary algebra that
(\ref{eq:system3}) is equivalent to

\begin{equation}
\begin{array}{l}
f_B=\sqrt[k]{1-\frac{\log f_A}{(f_A-1)ap}} \quad \textrm{if} \:\:0\leq f_A < 1;\:\: \forall f_B \:\: \textrm{if} \:\:f_A = 1 \\
 \\ f_A=1-\frac{1-\sqrt[k]{1-\frac{\log f_B}{(f_B-1)b}}}{p} \quad \textrm{if} \:\: 0\leq f_B < 1;\:\: \forall f_A \:\: \textrm{if} \:\: f_B =
 1.
\end{array}
\label{eq:f_ab_soln}
\end{equation}

\begin{figure}[t]
 \hspace{-0.5 cm}
\includegraphics[totalheight=0.25\textheight,
width=.55\textwidth]{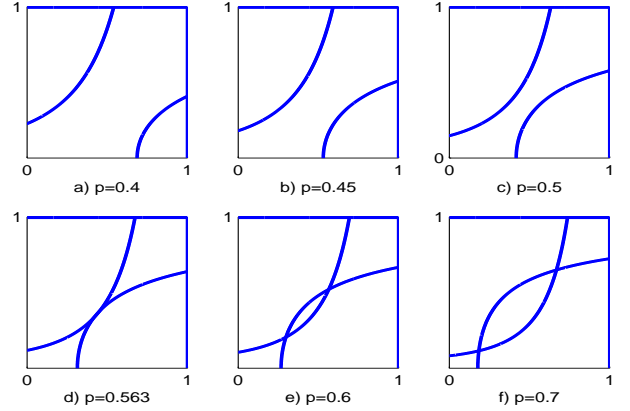} \caption{\sl Possible solutions of
the system (\ref{eq:f_ab_soln}) are depicted for several different
$p$ values when $a=b=3$ and $k=2$. In all figures, the $x$-axis
represents $f_A$ while the $y$-axis represents $f_B$. The critical
$p_c$ corresponds to the case where there is only one non-trivial
solution to the system, i.e., the case when the two curves are
tangential to each other. }
 \label{fig:solution_fab}
\end{figure}

In general, it may be difficult to derive an explicit expression
for $p_c$. Instead, we can solve (\ref{eq:f_ab_soln}) graphically
for a given set of parameters $a,b,k,p$ and infer the critical
threshold $p_c$ using numerical methods. For instance, Figure
\ref{fig:solution_fab} shows the possible solutions of the system
for several different $p$ values when $a=b=3$ and $k=2$. In
Figures \ref{fig:solution_fab}(a-c), we have $p<p_c$ and there is
only the trivial solution $f_A=f_B=1$ so that both networks go
into a complete fragmentation at steady state. In Figure
 \ref{fig:solution_fab}(d), we have $p=p_c$ and there exists one non-trivial
solution, since the two curves intersect tangentially at one
point. In Figures \ref{fig:solution_fab}(e-f), we have $p>p_c$ and
there exist two non-trivial intersection points corresponding to
two sets of giant component sizes. In these cases, the solution
corresponding to the cascading failures should be the point that
yields the larger giant component size. In other words, the
solution corresponds to the intersection point that is closer to
the  starting point of the iterative process (see
(\ref{eq:giant_comp})).

In the manner outlined above, we can find the critical threshold
$p_c$ for any fixed values of the parameters $a$, $b$ and $k$. As
illustrated in Figure \ref{fig:solution_fab}, we can further add
the tangential condition
\begin{equation}\label{eq:tangential}
\frac{df_A}{df_B} \times \frac{df_B}{df_A} = 1
\end{equation}
to the equations (\ref{eq:f_ab_soln}) since the critical $p_c$
value corresponds to the tangent point of the two curves given by
(\ref{eq:f_ab_soln}). Thus, the critical values $f_{A_c}$,
$f_{B_c}$ and $p_c$ can be computed (numerically) for any given
set of parameters through the following system of equations:
\begin{eqnarray}
f_{B} &=&\sqrt[k]{1-\frac{\log f_A}{(f_A-1)ap}} \quad \textrm{if}
\:\:0\leq f_A < 1; \label{eq:extra1}
\\
f_{A}&=&1-\frac{1-\sqrt[k]{1-\frac{\log f_B}{(f_B-1)b}}}{p} \quad
\textrm{if} \:\: 0\leq f_B < 1;\label{eq:extra2}\\
 &&~\frac{df_{A}}{df_B}\Huge{|}_{\textrm{Eq.} (\ref{eq:extra2})}
 \times \frac{df_{B}}{df_A}{|}_{\textrm{Eq.} (\ref{eq:extra1})} =
 1.
\label{eq:f_ab_soln2}
\end{eqnarray}

\begin{figure*}[!ht]
 \centering\subfigure[]
{\hspace{-0.5cm}
\includegraphics[totalheight=0.30\textheight,
width=0.48\textwidth]{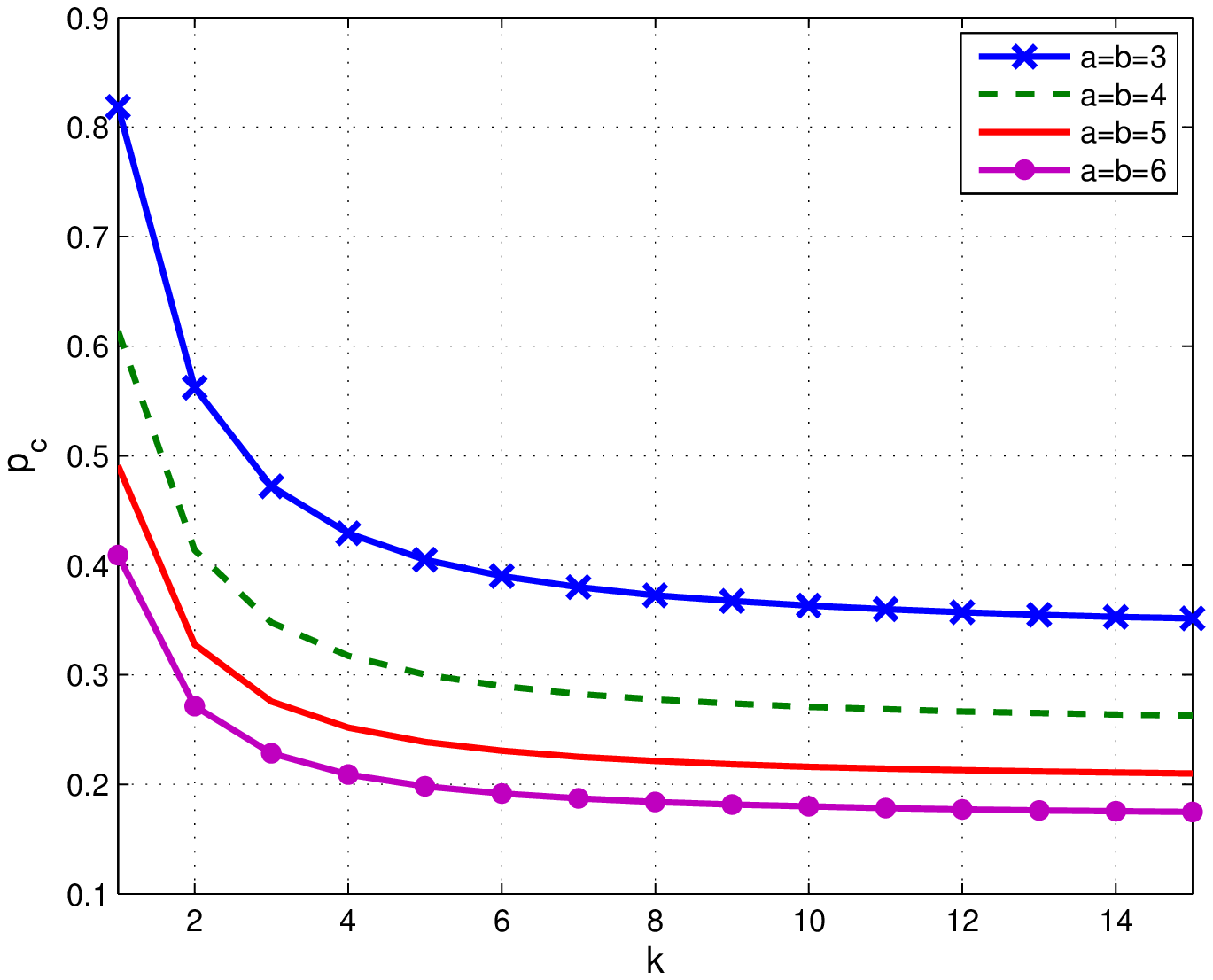} \label{fig:p_c} } \subfigure[]
{\hspace{-0.5cm}
\includegraphics[totalheight=0.30\textheight,
width=0.48\textwidth]{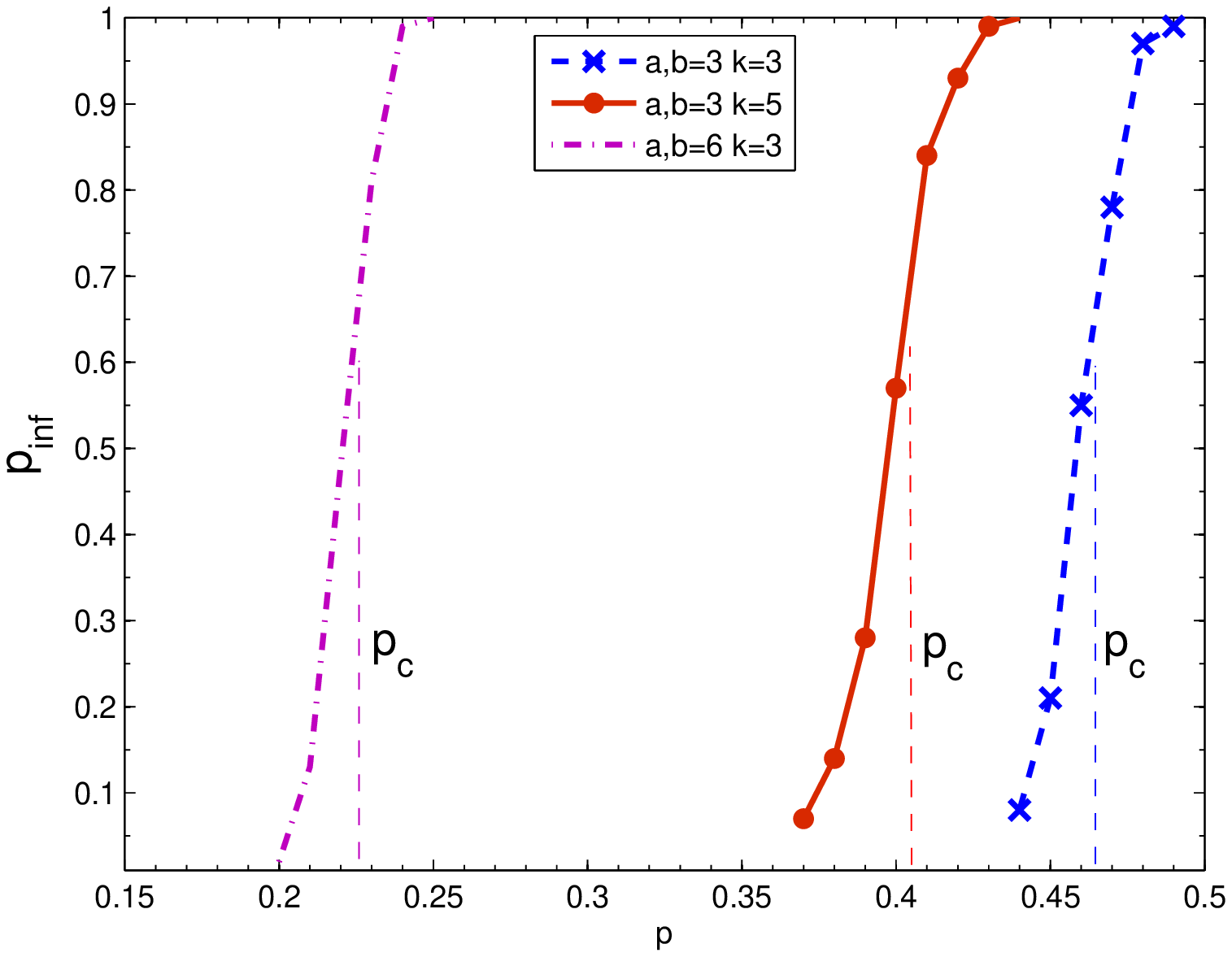} \label{fig:simu} } \caption{\sl{
$a)$ The critical $p_c$ value versus $k$ for the regular
allocation strategy (System $1$). The plots are obtained by
solving the system (\ref{eq:f_ab_soln2}) graphically for various
$a,b$ values. It can be seen that as $k$ increases the robustness
of the system increases and the critical fraction $p_c$ approaches
that of a single network; i.e., $\frac{1}{a}$ \cite{Bollobas}. b)
Experimental results for the regular allocation strategy (System
$1$) with $N=5000$ nodes. A fraction $1-p$ of the nodes are
randomly removed (from network $A$) and the corresponding
empirical probability $p_{\mbox{inf}}$ for the existence of a
functional giant component at steady state is plotted. As
expected, in all cases there is a sharp increase when $p$
approaches a critical threshold $p_c$; for ($a=b=3,k=3$),
($a=b=3,k=5$) and ($a=b=6,k=3$), the critical $p_c$ values are
roughly equal to $0.47$, $0.41$ and $0.23$, respectively. Clearly,
these $p_c$ values are in close agreement with the corresponding
ones of Figure \ref{fig:p_c} which are obtained analytically.}}
\end{figure*}

The analysis results are now corroborated by simulations. In
Figure \ref{fig:p_c}, we show the variation of $p_c$ with respect
to $k$ for different values of $a=b$, where the critical $p_c$
values are obtained by solving the system (\ref{eq:f_ab_soln2})
graphically. To verify these findings, we pick a few sets of
values $a$, $b$ and $k$ from the curves in Figure \ref{fig:p_c}
and run simulations with $N=5000$ nodes to estimate the
probability $p_{\mbox{inf}}$ of the existence of a functional
giant component in steady state. As expected \cite{Buldyrev}, in
all curves we see a sharp increase in $p_{\mbox{inf}}$ as $p$
approaches a critical threshold $p_c$. It is clear  that the
estimated $p_c$ values  from the sharp transitions in Figure
\ref{fig:simu} are in good agreement with the analysis results
given in Figure \ref{fig:p_c}.

\subsection{Numerical Results for System $2$}
\label{subsec:num2}

As in System $1$, the recursive process
(\ref{eq:recursive_sys_21})-(\ref{eq:recursive_sys_22}) of System
$2$ stops at an \lq\lq equilibrium point" where we have
$p'_{B2m-2}=p'_{B2m}=x$ and $p'_{A2m-1}=p'_{A2m+1}=y$. This yields
the transcendental equations
\begin{equation}
 \begin{array}{l}
x = p(1 - \sum\limits_{j=0}^{\infty} {\alpha _j (1 - P_B (y))^j }),\\
y = 1 - \sum\limits_{j=0}^{\infty} {\alpha _j (1 - pP_A (x))^j}.
  \end{array}
\label{eq:system2}
\end{equation}
The steady-state fraction of nodes in the giant components can be
computed by using the relations $\lim_{i \to \infty}p_{Ai} :=
P_{A_{\infty}}=xP_A(x)$ and $\lim_{i \to \infty}p_{Bi}
:=P_{B_{\infty}}=yP_B(y)$.

In particular, we assume that the inter-degree distribution $F$ at
each node is a Poisson distribution with mean $k$, and hence
\begin{equation}
\alpha _j  = e^{ - k} \frac{{k^j }}{{j!}}, \quad
j=0,1,2,...,\infty. \label{eq:alpha}
\end{equation}

Substituting (\ref{eq:P_A_B}) and (\ref{eq:alpha}) into
(\ref{eq:system2}), we get
\begin{eqnarray} \label{eq:system2_equal1}
x = p\left( {1 - \sum\limits_{j = 0}^\infty  {\frac{{k^j
}}{{j!}}e^{ - k}  {f_B}^j } } \right) = p\left( {1 - e^{ - k(1 -
f_B )} } \right),
\end{eqnarray}
and
\begin{eqnarray} \label{eq:system2_equal2}
y = 1 - \sum\limits_{j = 0}^\infty  {\frac{{k^j }}{{j!}}e^{ - k}
\left( {1 - p(1-f_A)} \right)^j } =   {1 - e^{ - k p (1 - f_A )} }
.
\end{eqnarray}
Next, putting (\ref{eq:system2_equal1}) and
(\ref{eq:system2_equal2}) into (\ref{eq:f_ab}), we find
\begin{equation}
\begin{array}{l}
f_A  = 1 + \frac{1}{{pk}}\ln \left( {1 + \frac{{\ln f_B
}}{{b\left( {1 - f_B } \right)}}} \right), \textrm{if} \:\:0\leq f_B < 1; \\
f_B  = 1 + \frac{1}{k}\ln \left( {1 + \frac{{\ln f_A }}{{ap\left(
{1 - f_A } \right)}}} \right), \textrm{if}
\:\:0\leq f_A < 1;\\
\forall f_A \:\: \textrm{if} \:\: f_B = 1;\quad \forall f_B \:\:
\textrm{if} \:\: f_A = 1.
\end{array}
\label{eq:f_ab_soln_sys2}
\end{equation}
As in the case for System $1$, the critical threshold $p_c$ for
System $2$ corresponds to the tangential point of the curves given
by (\ref{eq:f_ab_soln_sys2}), and can be obtained by solving
(\ref{eq:f_ab_soln_sys2}) graphically.
\begin{figure*}[!ht]
 \centering\subfigure[]
{\hspace{-0.5cm}
\includegraphics[totalheight=0.30\textheight,
width=0.48\textwidth]{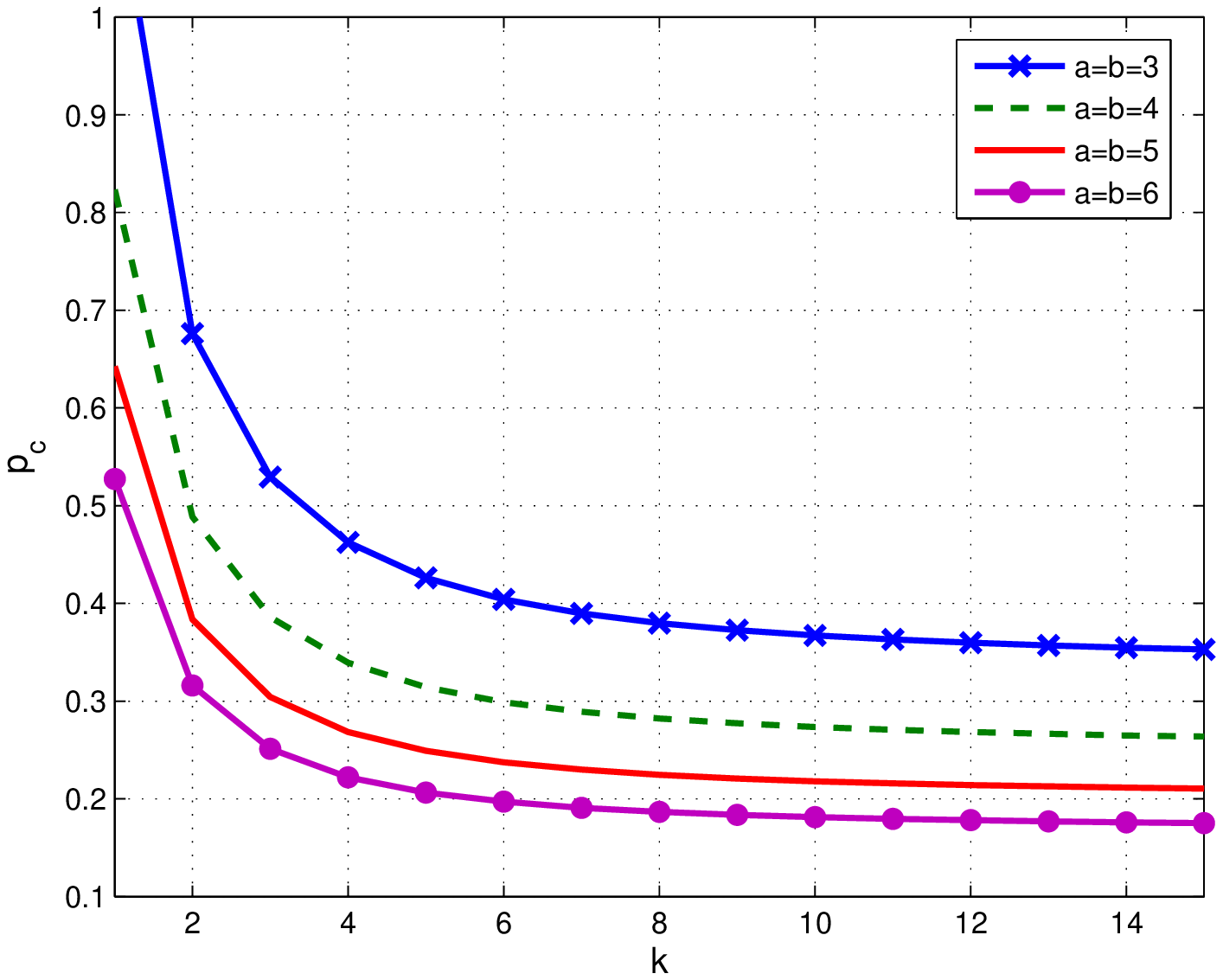} \label{fig:p_c_sys2} }
\subfigure[] {\hspace{-0.5cm}
\includegraphics[totalheight=0.30\textheight,
width=0.48\textwidth]{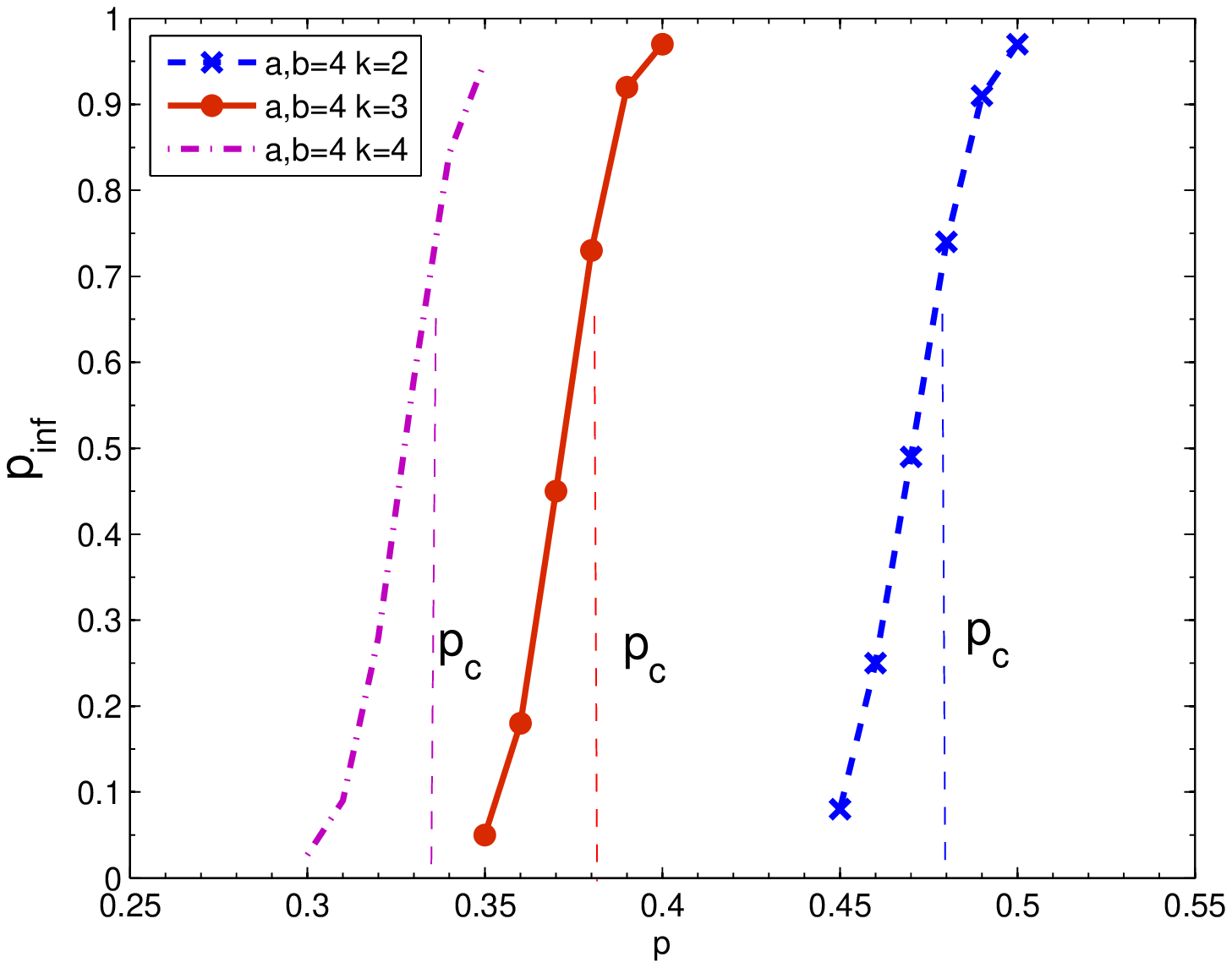} \label{fig:simu_sys2} }
\caption{\sl{ $a)$ The critical $p_c$ value versus $k$ for the
random allocation strategy (System $2$). The plots are obtained by
solving the system (\ref{eq:f_ab_soln_sys2}) graphically for
various $a,b$ values. It is seen that the critical $p_c$ can be
larger than one in some cases (e.g., for $a=b=3$ and $k=1$)
meaning that the system collapses already without any node being
attacked. This is because, due to the random allocation of
inter-edges, a non-negligible fraction of the nodes receive no
inter-edges and become automatically non-functional even if they
are not attacked. $b)$ Experimental results for System $2$ with
$N=5000$ nodes. A fraction $1-p$ of the nodes are randomly removed
(from network $A$) and the corresponding empirical probability
$p_{\mbox{inf}}$ for the existence of a functional giant component
at the steady state is plotted. As expected, in all cases there is
a sharp increase when $p$ approaches to a critical threshold
$p_c$; for ($a=b=4,k=2$), ($a=b=4,k=3$) and ($a=b=k=4$), the
critical $p_c$ values are roughly equal to $0.480$, $0.380$ and
$0.335$, respectively. Clearly, these $p_c$ values are in close
agreement with the corresponding ones of Figure \ref{fig:p_c_sys2}
which are obtained analytically.}}
\end{figure*}

We now check the validity of these analytical results via
simulations. In Figure \ref{fig:p_c_sys2}, we show the variation
of analytically obtained $p_c$ values with respect to average
inter-degree $k$ for different values of $a=b$. To verify these
results, we pick a few sets of values $a$, $b$ and $k$ from the
curves in Figure \ref{fig:p_c_sys2} and run simulations with
$N=5000$ nodes to estimate the probability $p_{\mbox{inf}}$ of the
existence of a functional giant component in steady state. As
expected \cite{Buldyrev}, in all curves we see a sharp increase in
$p_{\mbox{inf}}$ as $p$ approaches a critical threshold $p_c$. It
is also clear from Figure \ref{fig:simu_sys2} that, for all
parameter sets, such sharp transition occurs when $p$ is close to
the corresponding $p_c$ value given in Figure \ref{fig:p_c_sys2}.

\subsection{A Comparison of System Robustness}

\begin{figure*}[!ht]
\label{fig:compare} \centering\subfigure[]{\hspace{-0.5cm}
\includegraphics[totalheight=0.3\textheight,
width=0.5\textwidth]{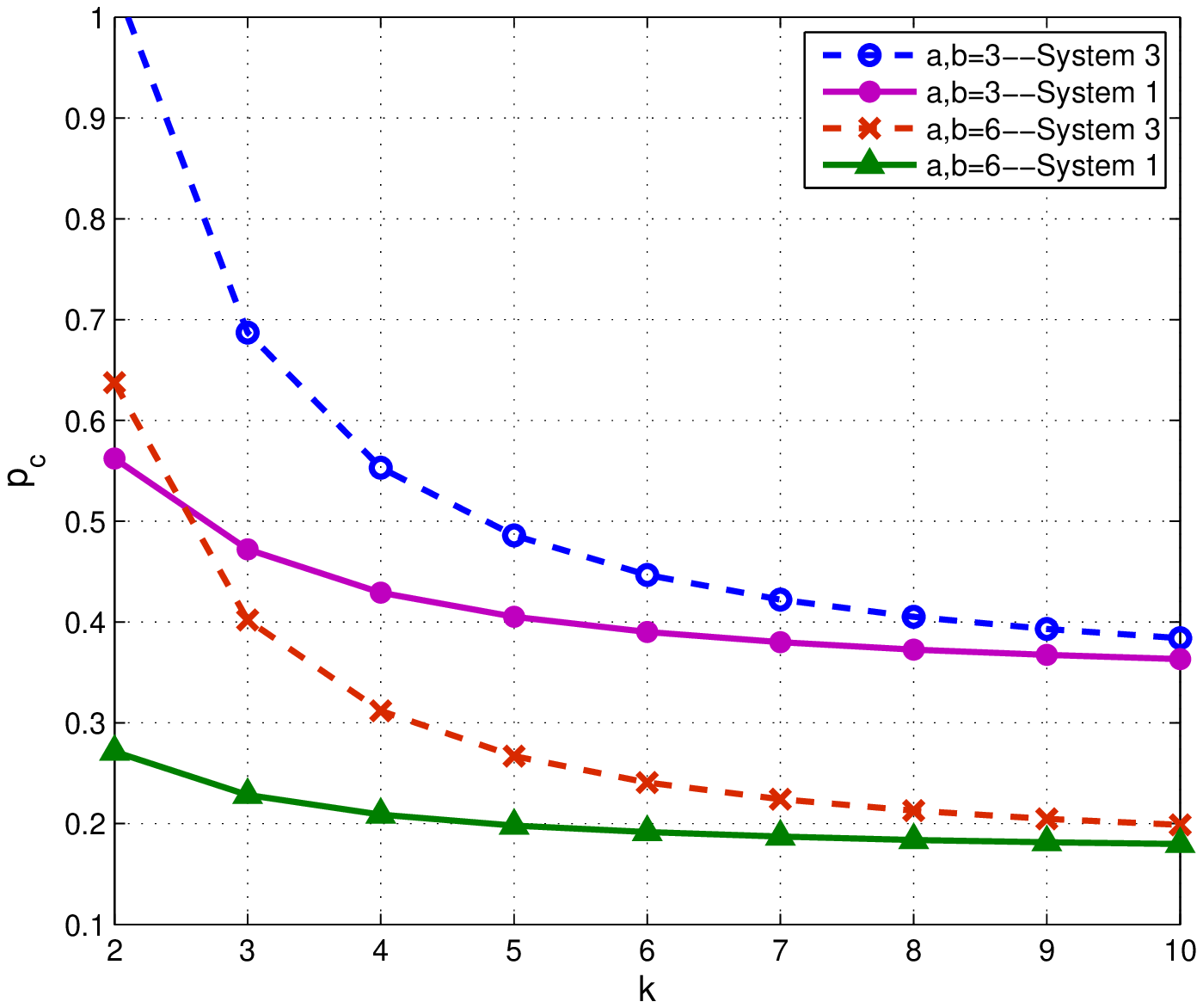} \label{fig:reg_vs_binom} }
\subfigure[] {\hspace{-0.5cm}
\includegraphics[totalheight=0.3\textheight,
width=0.5\textwidth]{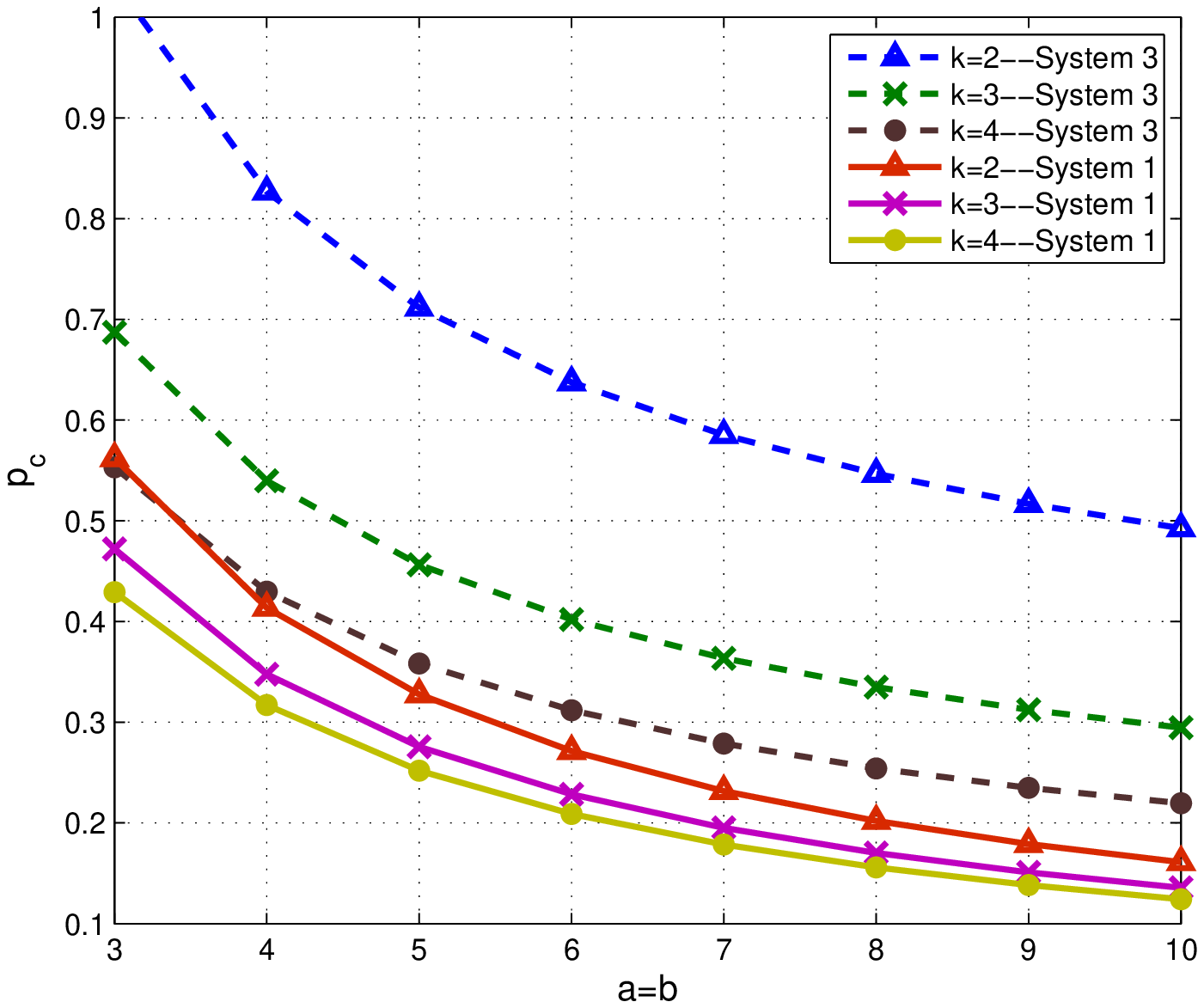} \label{fig:reg_vs_binom2} }
\subfigure[]{\hspace{-0.5cm}
\includegraphics[totalheight=0.3\textheight,
width=0.5\textwidth]{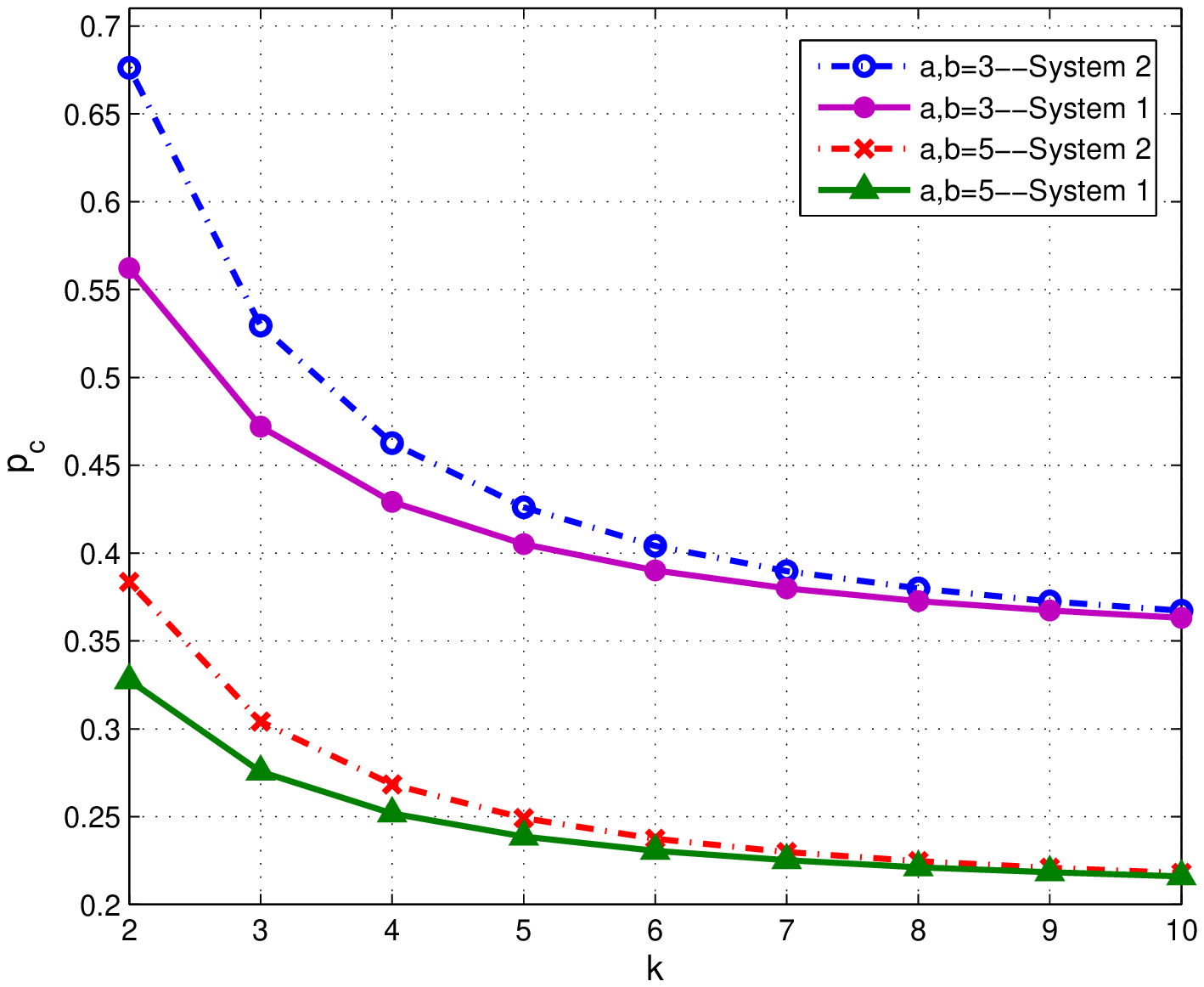} \label{fig:reg_vs_binom3} }
\subfigure[] {\hspace{-0.5cm}
\includegraphics[totalheight=0.3\textheight,
width=0.5\textwidth]{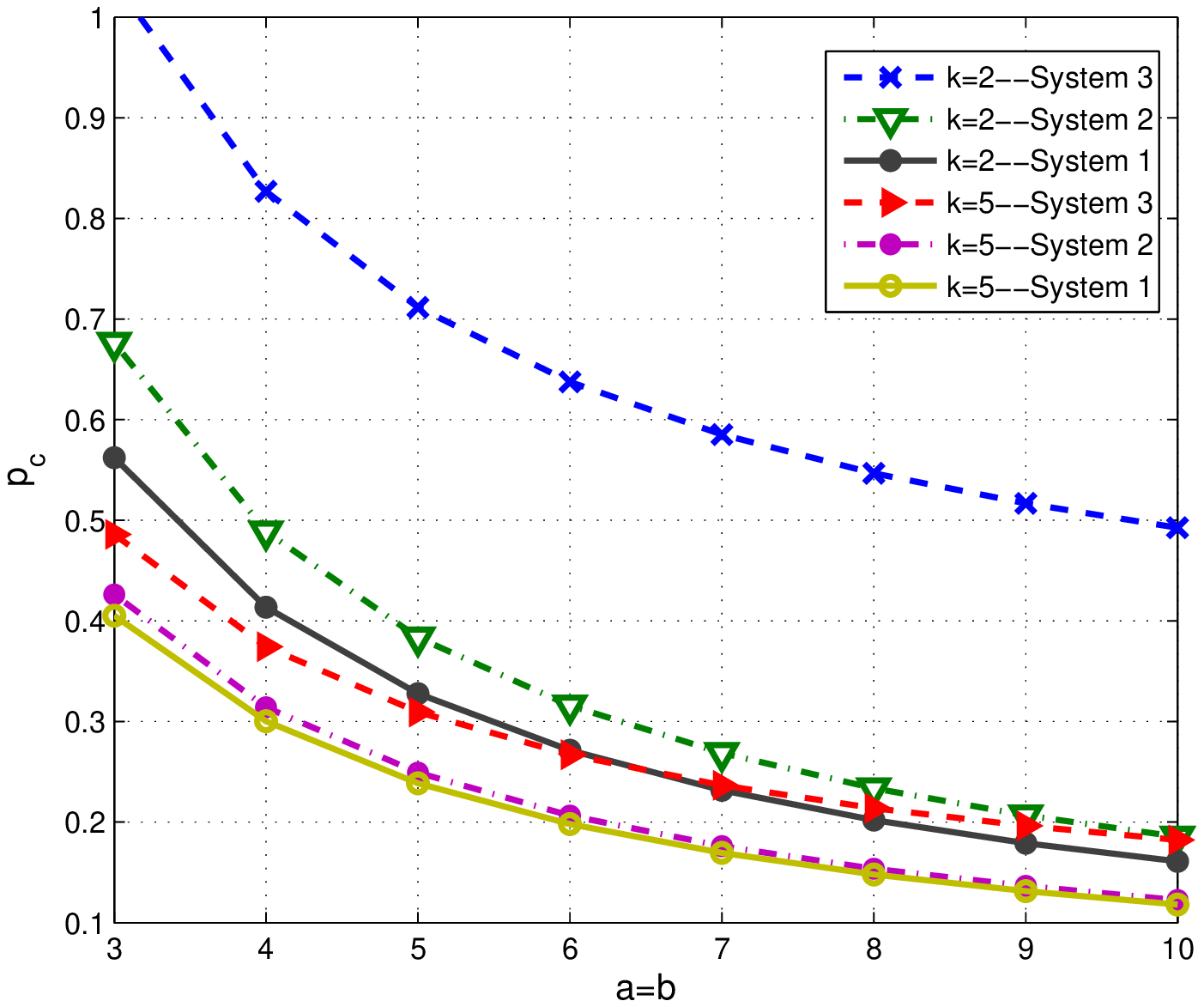} \label{fig:reg_vs_binom4} }
\caption{\sl{ A comparison of System $1$, System $2$ and System
$3$ in terms of their critical $p_c$ values when expected
inter-degree of any node is set to $k$. For System $2$ and System
$3$, the distribution of the number of inter-edges is assumed to
be Poisson. In all figures, dashed lines correspond to System $3$,
dash-dot lines represent System $2$, and solid lines stand for
System $1$. $a)$ $p_c$ v.s. $k$ is depicted for different values
of $a=b$ in System $1$ and System $3$. $b)$ $p_c$ v.s. $a=b$ is
depicted for various $k$ values in System $1$ and System $3$. In
all cases, we see that the regular allocation of bi-directional
inter-edges yields a smaller $p_c$ than the Poisson distribution
of unidirectional inter-edges with the same mean value $k$. $c)$
$p_c$ v.s. $k$ is depicted for different values of $a=b$ in System
$1$ and System $2$. It is clear that System $1$ yields a lower
$p_c$ (and thus a higher robustness) than System $2$ in all cases.
$d)$ $p_c$ v.s. $a=b$ is depicted for various $k$ values in System
$1$, System $2$ and System $3$. In all cases System $1$ yields the
lowest $p_c$ (i.e., highest robustness), while System $3$ has the
highest $p_c$ (i.e., lowest robustness) and System $2$ stands in
between.}}
\end{figure*}

In Section \ref{subsec:reg_vs_rand}  and \ref{subsec:bi_uni}, we
have analytically proved that the regular allocation of
bi-directional inter-edges leads to the strongest robustness
against random attacks. To get a more concrete sense, we now
numerically compare the system robustness of these strategies in
terms of their critical thresholds $p_c$. Specifically, we
consider coupled Erd\H{o}s-R\'enyi networks with mean
intra-degrees $a$ and $b$. For the sake of fair comparison, we
assume that the mean inter-degree is set to $k$ for all systems;
in both Systems $2$ and $3$, the inter-degree distribution $F$ at
each node is assumed to be Poisson. The critical threshold value
$p_c$ corresponding to all three strategies are compared under a
variety of conditions. For Systems $1$ and $2$, we use the
numerical results derived in Section \ref{subsec:num1} and Section
\ref{subsec:num2}, respectively, while for System $3$ we use the
numerical results provided in \cite{ShaoBuldyrevHavlinStanley}.

First, we compare System $1$ with System $3$ to see the difference
between the proposed regular inter-edge allocation strategy and
the strategy  in \cite{ShaoBuldyrevHavlinStanley}. Figure
\ref{fig:reg_vs_binom} depicts $p_c$ as a function of mean
inter-degree $k$ for various values of $a=b$, while Figure
\ref{fig:reg_vs_binom2} depicts the variation of $p_c$ with
respect to $a=b$ for different $k$ values. In all cases, it is
seen that regular allocation of bi-directional inter-edges yields
a much smaller $p_c$ (and thus, a more robust system) than random
allocation of unidirectional inter-edges. For instance, for
$a=b=k=4$, System $3$ \cite[Figure 2]{ShaoBuldyrevHavlinStanley}
gives $p_c=0.43$, whereas, as seen via
Figure~\ref{fig:reg_vs_binom}, System $1$ yields a critical
threshold at $0.317$. This is a significant difference since it
means that System $3$ can have a functioning giant component
despite a random failure of at most $57 \%$ of the nodes, whereas
System $1$, which uses the regular inter-edge allocation scheme
proposed in this paper, is resistant to a random failure of up to
$68\%$ of the nodes. Indeed, in some cases, our strategy can
outperform that in \cite{ShaoBuldyrevHavlinStanley} even with half
the (mean) inter-degree per node. For instance, when $a=b=4$, our
strategy yields $p_c = 0.414$ with only $k=2$ as compared to
$p_c=0.43$ of the System $3$ with $k=4$.

We also compare System $1$ with System $2$ in order to see the
improvement in allocating bi-directional edges regularly rather
than randomly. Figure \ref{fig:reg_vs_binom3} depicts $p_c$ as a
function of mean inter-degree $k$ for various values of $a=b$. It
is seen that, in all cases, System $1$ yields a lower $p_c$ (and
thus a more robust system) than System $2$. For example, when
$a=b=3$ and $k=2$, we get $p_{c}=0.56$ for System $1$, while for
System $2$, we find that $p_c=0.68$. The difference is significant
in that it corresponds to a resiliency against a random failure of
up to $44\%$ of the nodes in System $1$ as compared to $32\%$ in
System $2$.

Finally, in order to better illustrate the optimality of System
$1$ in terms of system robustness, we depict in Figure
\ref{fig:reg_vs_binom4} the variation of $p_c$ with respect to
$a=b$ for different values of $k$ in all three systems. It is
clear that the proposed regular allocation strategy in System $1$
always yields the lowest $p_c$ and thus provides the best
resiliency against random attacks. We also see that System $2$
always outperforms System $3$, showing the superiority (in terms
of robustness) of using bi-directional inter-edges rather than
unidirectional edges.

We believe that the drastic improvement in robustness against
random attacks seen in System $1$ has its roots as follows. First,
in the absence of intra-topology information, it is difficult to
tell which nodes play more important roles in preserving the
connectivity of the networks. Thus, in order to combat {\em
random} attacks, it is reasonable to treat all nodes equally and
give them equal priority in inter-edge allocation. Secondly, in
Systems $2$ and $3$, there may exist a non-negligible fraction of
nodes with no inter-edge support from the other network. Those
nodes are automatically non-functional even if they are not
attacked. But, the regular allocation scheme promises a guaranteed
level of support, in terms of inter-edges, for all nodes in both
networks. Finally, using bi-directional inter-edges ensures that
the amount of support provided is equal to the amount of support
being received for each node. Thus, the use of bi-directional
inter-edges increases the {\em regularity} of the
support-dependency relationship relative to unidirectional
inter-edges, and this may help improve the system robustness.

\section{Conclusion and Future Work}
\label{sec:conclusion}

We study the robustness of a cyber-physical system in which a
cyber-network overlays a physical-network. To improve network
robustness against random node failures, we develop and study a
regular allocation strategy that allots a fixed number of
inter-network edges to each node. Our findings reveal that the
proposed regular allocation strategy yields the optimal robustness
amongst all strategies when no information regarding the
intra-topologies of the individual networks is available. For
future work, we conjecture that in the presence of such
information, the topology of the networks can be exploited to
further improve the robustness of cyber-physical systems against
cascading failures.

It is also of interest to study models that are more realistic
than the existing ones. For instance, in a realistic setting, one
can expect to see a certain correlation between the inter-edges
and the intra-edges of a system owing to the geographical
locations of the nodes. Also, some of the
nodes may be {\em autonomous}, meaning
that they do not depend on nodes of the other network to function
properly; in that case, one can expect the regular allocation strategy 
to still be the optimum if the nodes that are autonomous are not known.
Clearly, there are still many open
questions centered around network interdependence in
cyber-physical systems. We are currently investigating related
issues along this avenue.

\section{Acknowledgments}
We thank the anonymous reviewers for their careful
reading of the original manuscript;
their comments helped improve the final version of this paper.
We also thank Prof. Armand Makowski for his insightful comments. Part
of this material was presented in \cite{YaganQianZhangCochran}.
This research was supported in part by the U.S. National Science
Foundation grants No. CNS-0905603, CNS-0917087, and the  DTRA
grant HDTRA1-09-1-0032.

\bibliographystyle{unsrt}

\begin{thebibliography}{1}

\bibitem{BarabasiAlbert}
A. L. Barab\'{a}si and L. Albert, \lq\lq Emergence of Scaling in
Random Networks," \sl{Science} {\bf 286}:509-512, 1999.


\bibitem{Bollobas}
B. Bollob\'{a}s, {\sl Random Graphs},
Cambridge Studies in Advanced Mathematics, Cambridge University
Press, Cambridge (UK), 2001.

\bibitem{Buldyrev}
S.V. Buldyrev, R.~Parshani, G.~Paul, H.E. Stanley, and S.~Havlin,
\lq\lq Catastrophic cascade of failures in interdependent
networks," \sl {Nature}, {\bf 464}:1025--1028, 2010.

\bibitem{BuldyrevShereCwilich}
S. V. Buldyrev, N. W. Shere, and G. A. Cwilich, \lq\lq
Interdependent networks with identical degrees of mutually
dependent nodes," {\sl Physical Review E} {\bf 83}:016112, 2011.

\bibitem{callaway2000network}
D.S. Callaway, M.E.J. Newman, S.H. Strogatz, and D.J. Watts \lq\lq
Network robustness and fragility: Percolation on random graphs,"
\sl{Physical Review Letters}, {\bf 85(25)}:5468--5471, 2000.

\bibitem{ChoGohKim}
W. Cho, K.I. Goh and I.M. Kim, \lq\lq Correlated couplings and
robustness of coupled networks," Available online at
\url{arXiv:1010.4971v1 [physics.data-an]}.

\bibitem{cohen2000resilience}
R.~Cohen, K.~Erez, D.~Ben-Avraham, and S.~Havlin, \lq\lq
Resilience of the internet to random breakdowns," \sl{Physical
Review Letters}, {\bf 85(21)}:4626--4628, 2000.

\bibitem{CohenHavlin}
R. Cohen and S. Havlin, {\sl Complex networks: structure,
robustness and function}, Cambridge University Press, United
Kingdom, 2010.

\bibitem{CPS}
CPS~Steering Group, \lq\lq Cyber-physical systems executive
summary, 2008," Available online at
\url{http://varma.ece.cmu.edu/summit/CPS-Executive-Summary.pdf}.

\bibitem{HuangGaoBuldyrevHavlinStanley}
X. Huang, J. Gao, S. V. Buldyrev, S. Havlin, and H. E. Stanley,
\lq \lq Robustness of interdependent networks under targeted
attack", \sl{Physical Review E} {\bf 83}: 065101, 2011.

\bibitem{knuth1981art}
D. E. Knuth, {\sl The Art of Computer Programming}, Volume 2,
Addison--Wesley, 1981.

\bibitem{newman2002spread}
M.E.J. Newman, \lq \lq Spread of epidemic disease on networks,"
\sl{Physical Review E} {\bf 66(1)}:16128, 2002.

\bibitem{newman2001random}
M.E.J. Newman, S.H. Strogatz, and D.J. Watts. \lq\lq Random graphs
with arbitrary degree distributions and their applications," \sl{
Physical Review E} {\bf 64(2):}26118, 2001.

\bibitem{ParshaniBuldyrevHavlin}
R. Parshani, S. V. Buldyrev, and S. Havlin, \lq\lq Interdependent
Networks: Reducing the Coupling Strength Leads to a Change from a
First to Second Order Percolation Transition," \sl{Physical Review
Letters} {\bf 105}:048701, 2010.


\bibitem{SchneiderArajuoHavlinHerrman}
C. M. Schneider, N. A. M. Araujo, S. Havlin and H. J. Herrmann,
\lq \lq Towards designing robust coupled networks," Available
online at \url{arXiv:1106.3234v1 [cond-mat.stat-mech]}.

\bibitem{ShaoBuldyrevHavlinStanley}
J.~Shao, S.V. Buldyrev, S.~Havlin, and H.E. Stanley, \lq \lq
Cascade of failures in coupled network systems with multiple
  support-dependent relations,"
\sl{Physical Review E} {\bf 83}:036116, 2011.

\bibitem{Vespignani}
A. Vespignani, \lq \lq Complex networks: The fragility of
interdependency," \sl{Nature} {\bf 464}: 984-985, April 2010.

\bibitem{YaganQianZhangCochran}
O. Ya\u{g}an, D. Qian, J. Zhang, and D. Cochran, \lq\lq  On
allocating interconnecting links against cascading failures in
cyber-physical networks," Proceedings of the Third International
Workshop on Network Science for Communication Networks, (NetSciCom
2011), April 2011.

\end{thebibliography}

\begin{IEEEbiography}[{\includegraphics[width=1.0in,height =1.5in,clip,keepaspectratio]
{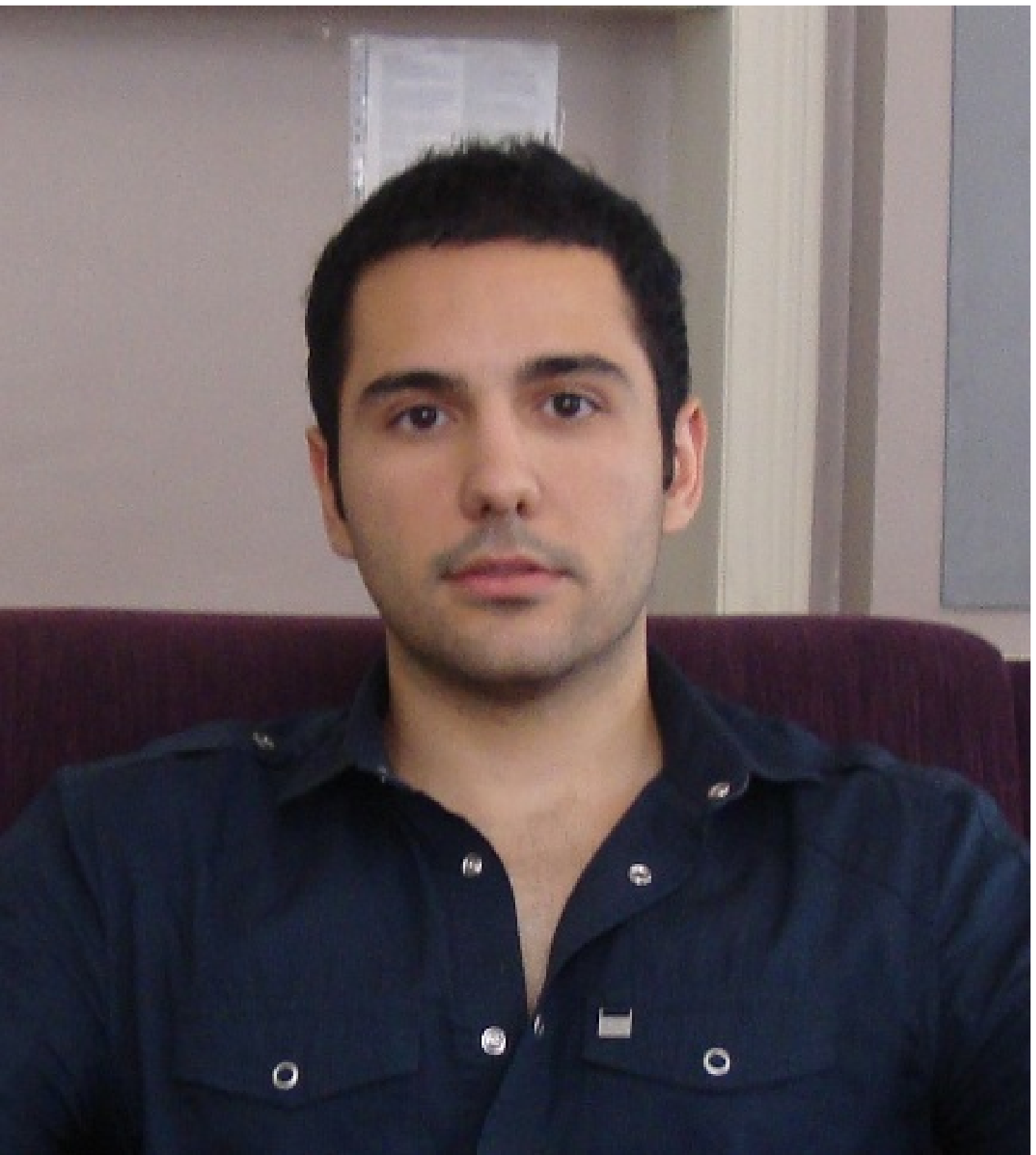}}]
{Osman Ya\u{g}an} (S'07) received the B.S.
degree in Electrical and Electronics Engineering from the Middle
East Technical University, Ankara (Turkey) in 2007, and the Ph.D.
degree in Electrical and Computer Engineering from the University
of Maryland, College Park, MD in 2011.

He was a visiting Postdoctoral Scholar at Arizona State University 
during Fall 2011. Since December 2011, he has been a Postdoctoral Research
Fellow in the Cyber Security Laboratory (CyLab)
at the Carnegie Mellon University. His research interests
include wireless network security, 
dynamical processes in complex networks, 
percolation theory, random
graphs and their applications.
\end{IEEEbiography}

\begin{IEEEbiography}[{\includegraphics[width=.9in,height =1.2in,clip,keepaspectratio]
{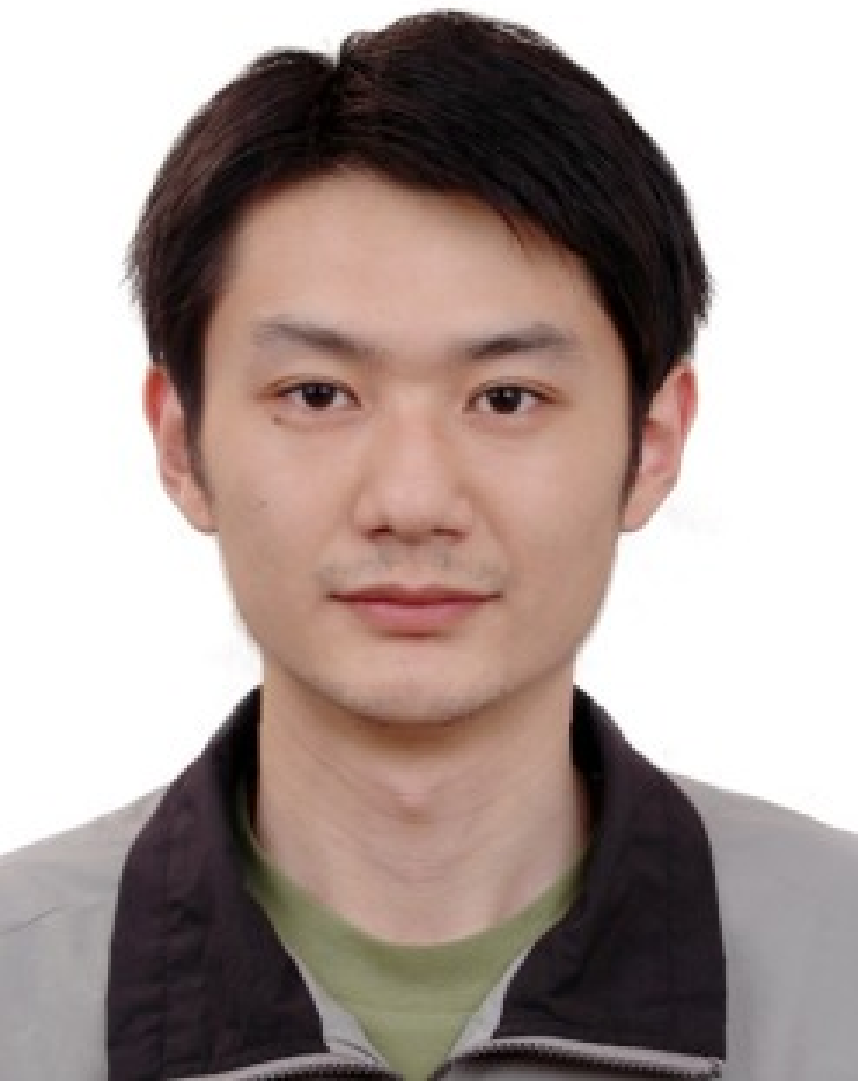}}]
{Dajun Qian} received his B.S. and M.S. degrees of Electrical Engineering from 
Southeast University, Nanjing, China, in 2006 and 2008, respectively. 
Currently, he is a Ph.D. student in the Department of Electrical, Computer and Energy Engineering, 
Arizona State University, Tempe, AZ. 
His research interests include wireless communications, social networks and cyber-physical systems.
\end{IEEEbiography}

\begin{IEEEbiography}[{\includegraphics[width=1in,height =1.25in,clip,keepaspectratio]
{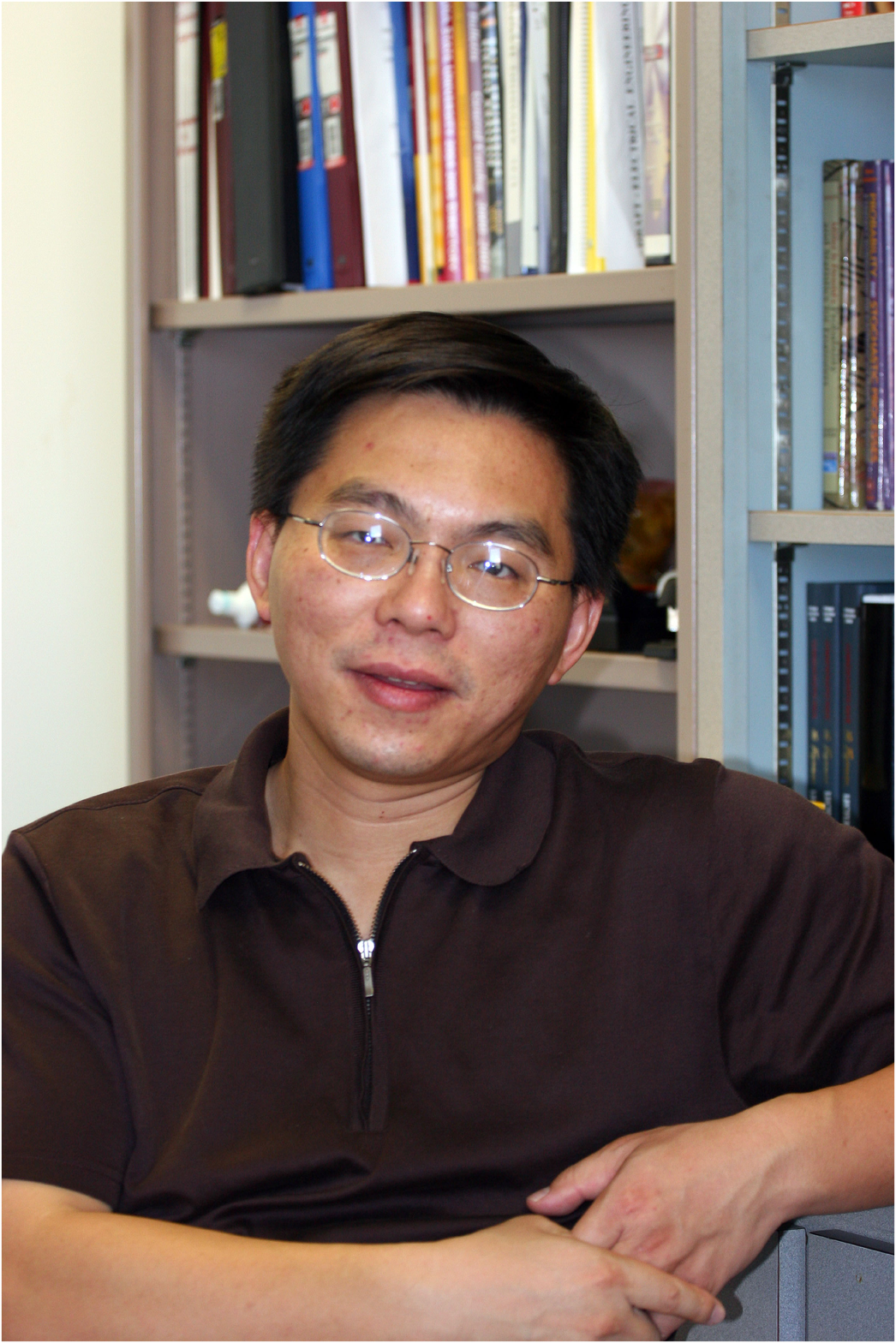}}]
{Junshan Zhang} (S'98-M'00-F'12) received his Ph.D. degree from the School of ECE at Purdue University in
2000. He joined the EE Department at Arizona State University in August 2000, where he
has been Professor since 2010. His research interests include communications networks,
cyber-physical systems with applications to smart grid, stochastic modeling and analysis, and
wireless communications. His current research focuses on fundamental problems in information
networks and network science, including network optimization/control, smart grid, cognitive
radio, and network information theory.

Prof. Zhang is a fellow of the IEEE, and a recipient of the ONR Young Investigator Award in
2005 and the NSF CAREER award in 2003. He received the Outstanding Research Award from
the IEEE Phoenix Section in 2003. He served as TPC co-chair for WICON 2008 and IPCCC'06,
TPC vice chair for ICCCN'06, and the general chair for IEEE Communication Theory Workshop
2007. He was an Associate Editor for IEEE Transactions on Wireless Communications. He
is currently an editor for the Computer Network journal and IEEE Wireless Communication
Magazine. He co-authored a paper that won IEEE ICC 2008 best paper award, and one of his
papers was selected as the INFOCOM 2009 Best Paper Award Runner-up. He is TPC co-chair
for INFOCOM 2012.
\end{IEEEbiography}

\begin{IEEEbiography}[{\includegraphics[width=1in,height =1.25in,clip,keepaspectratio]{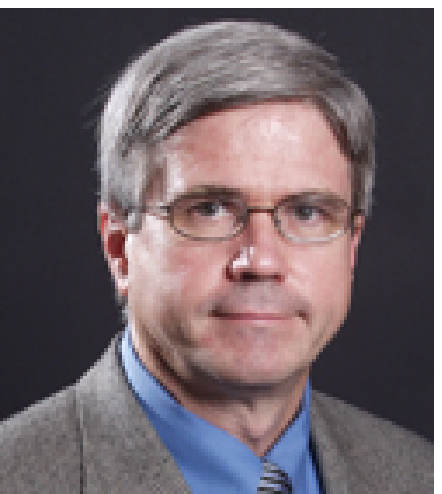}}]
{Douglas Cochran} (S'86-A'90-M'92-SM'96) holds M.S. and Ph.D. degrees in applied 
mathematics from Harvard University, Cambridge, MA, and degrees in mathematics from 
the Massachusetts Institute of Technology, Cambridge, and the University of California, San Diego.

Since 1989, he has been on the faculty of the School of Electrical, Computer and Energy Engineering, 
Arizona State University (ASU), Tempe, and is also affiliated with the School of Mathematical 
and Statistical Sciences. Between 2005 and 2008, he served as Assistant Dean for Research in the Ira A. 
Fulton School of Engineering at ASU. Between 2000 and 2005, 
he was Program Manager for Mathematics at the Defense Advanced Research 
Projects Agency (DARPA) and he held a similar position in the U.S. Air Force 
Office of Scientific Research between 2008 and 2010. Prior to joining the ASU 
faculty, he was Senior Scientist at BBN Systems and Technologies, Inc., during 
which time he served as resident scientist at the DARPA Acoustic Research Center and the Naval Ocean Systems Center. 
He has been a visiting scientist at the Australian Defence Science and Technology Organisation and served as 
a consultant to several technology companies.

Prof. Cochran was General Co-Chair of the 1999 IEEE International Conference on 
Acoustics, Speech and Signal Processing (ICASSP-99) and Co-Chair of the 1997 U.S.-Australia 
Workshop on Defense Signal Processing. He has also served as Associate Editor for book series and journals, 
including the IEEE TRANSACTIONS ON SIGNAL PROCESSING. 
His research is in applied harmonic analysis and statistical signal processing.
\end{IEEEbiography}

\end{document}